\documentclass[]{aa} 
\usepackage{graphicx,txfonts,amssymb,natbib,epsfig}
\authorrunning{M. Meneghetti et al.}
\titlerunning{Realistic simulations of lensing by galaxy clusters} 

\begin{document}

\title{Realistic simulations of gravitational lensing by galaxy clusters: extracting arc parameters from mock DUNE images}

 \author{Massimo Meneghetti\inst{1,2}, Peter Melchior\inst{2}, Andrea
  Grazian\inst{3},  Gabriella De Lucia \inst{4}, Klaus Dolag\inst{4}, Matthias Bartelmann\inst{2}, Catherine Heymans\inst{5,9}, Lauro Moscardini\inst{6,7}, Mario Radovich\inst{8}}

\institute {$^1$ INAF-Osservatorio
  Astronomico di Bologna, Via Ranzani 1, 40127 Bologna, Italy \\ $^2$ ITA, Zentrum f\"ur Astronomie, Universit\"at
  Heidelberg, Albert \"Uberle Str. 2, 69120 Heidelberg, Germany \\ $^3$
  INAF-Osservatorio Astronomico di Roma, Via di Frascati 33, 00040
  Roma, Italy \\ $^4$ Max-Planck-Institut f\"ur
  Astrophysik, Karl-Schwarzschild-Str. 1 85748, Garching bei Muenchen,
  Germany \\ $^5$ Department of Physics and Astronomy, The University of British Columbia, 6224 Agricultural Road, Vancouver, V6T 1Z1 \\$^6$ Dipartimento di Astronomia, Universit\`a di Bologna,
  Via Ranzani 1, 40127 Bologna, Italy \\ $^7$ INFN-National Institute for Nuclear Physics, Sezione di Bologna, Viale Berti Pichat 6/2. 40127, Bologna \\
  $^8$ INAF - Osservatorio Astronomico di Capodimonte, Salita Moiarello 16, 80131, Napoli \\
  $^9$ Institut d'Astrophysique de Paris, UMR7095 CNRS, 98 bis bd Arago, 75014,
Paris}

\date{\emph{Astronomy \& Astrophysics, submitted}}

\abstract {} {We present a newly developed code that allows
  simulations of optical observations of galaxy fields with a variety
  of instruments. The code incorporates gravitational lensing effects
  and is targetted at simulating lensing by galaxy clusters. Our goal
  is to create the tools required for comparing theoretical
  expectations with observations to obtain a better understanding
  of how observational noise affects lensing applications such as mass
  estimates, studies on the internal properties of galaxy clusters and
  arc statistics.}  {Starting from a set of input parameters,
  characterizing both the instruments and the observational
  conditions, the simulator provides a virtual observation of a patch
  of the sky. It includes several sources of noise such as photon-noise,
  sky background, seeing, and instrumental noise. Ray-tracing through
  simulated mass distributions accounts for gravitational
  lensing. Source morphologies are realistically simulated based on
  shapelet decompositions of galaxy images retrieved from the
  GOODS-ACS archive. According to their morphological class,
  spectral-energy-distributions are assigned to the source galaxies in
  order to reproduce observations of each galaxy in arbitrary
  photometric bands.} {We illustrate our techniques showing virtual
  observations of a galaxy-cluster core as it would be observed with
  the space telescope DUNE, which was recently proposed to ESA within
  its "Cosmic vision" programme. We analyze the simulated images
  using methods applicable to real observations and measure the
  properties of gravitational arcs. In particular, we focus on the
  determination of their length, width and curvature radius.} {We find
  that arc properties strongly depend on several properties of the
  sources. In particular, our results show that compact, faint or low
  surface brightness galaxies that are barely detectable are more
  easily distorted as arcs with large length-to-width ratios. We
  conclude that realistic lensing simulations can be obtained with the
  method proposed here. They will be essential for evaluating and
  improving analysis techniques currently used for cosmological
  interpretations of cluster lensing.}


\maketitle

\section{Introduction}
Gravitational lensing of distant galaxies is one of the most powerful
tools for investigating the content of the universe and for
understanding how the matter is distributed within the cosmic
structures.

Being the most massive objects in the universe, galaxy clusters are
the most powerful existing gravitational lenses. This makes them
suitable for a large number of purposes.

Clusters are able to magnify distant galaxies. Thus, they are natural
gravitational telescopes, that can be used to investigate the first
sources of light in the universe
\citep{EL01.1,SCHA03.1,KN04.1,LI07.1}.

Strong lensing features, like gravitational arcs and multiple images
are observed in the central, most dense regions of the lenses
\citep{FO88.1,LY89.1}. Hence, they can be used to constrain the inner
structure of clusters
\cite[e.g][]{LI07.1,CO05.1,OG03.1,SA03.1,ME07.1,ME07.2}.

Gravitational arcs are also used in a variety of cosmological
applications.  Their frequency on the sky reflects the abundance, the
concentration and the redshift distribution of massive lenses in the
universe. Moreover, it has been shown that the probability of forming
arcs is enhanced during cluster mergers \citep{TO04.1,FE05.1}. Thus,
arc statistics are a powerful tool for tracing structure formation
and for constraining the cosmological parameters
\citep{BA98.2,LI05.1,ME05.2}.

Moreover, since they form along critical lines, whose size scales with
the source and lens redshifts, lensed images arising from multiple
sources at different redshifts can be used to constrain the geometry
of the universe \citep{GO02.2,ME07.2}, and thus also cosmological
parameters.

In the outer regions of clusters, the distortions induced by the
cluster potentials are weaker. Nevertheless, using suitable algorithms
\cite[e.g.][and references therein]{BA01.1}, the distortion field can
be inverted to obtain two dimensional surface density maps of the the
lenses \cite[see][for some recent
examples]{CL04.1,ST06.1,PH07.1}. Strong and weak lensing can also be
used jointly to improve the mass reconstrution in a non-parametric way
\citep{BR04.1,CA06.1,DI07.1}.

Although the potential power of gravitational lensing for the above
mentioned applications has been already demonstrated, different
implementations of methods adopted for extracting the desired
information from the lensing data often give contradictory
results. An example is given by the so called "arc statistics
problem", i.e. the claimed inconsistency between the observed and the
predicted number of highly distorted gravitational arcs expected to be
seen on the sky in the standard $\Lambda$CDM cosmology
\citep{BA98.2,DA04.1,WA03.1,LU99.1,GL03.1,ZA03.1,LI05.1}. Moreover,
testing a particular analysis technique and/or comparing the results
to the theoretical expectations is difficult because observational
noise introduces uncertainties that need to be properly modeled
\citep[see e.g.][]{RH07.1}.

Following an approach that turned out to be extremely successful in
cosmic shear studies \cite[see e.g][]{HEY06.1,MAS07.2}, we aim at
overcoming these difficulties and facilitating the comparison between
theory and observations by performing realistic simulations, including
the relevant observational effects. Using suitable ray-tracing codes,
the gravitational lensing produced by realistic mass distributions
obtained from numerical N-body and hydrodynamical simulations can be
included. Simulated images can finally be analyzed like real
observations.

In this paper, we describe a simulation pipeline that we will use to
simulate observations of lensing events in galaxy clusters, but which
can easily be used also for simulating lensing by large scale
structures in wide fields. These simulations will be used in
forthcoming papers to evaluate the accuracy of several lensing
techniques mentioned above.

The plan of the paper is as follows. In Sect.~\ref{sect:simulator} we
present the architecture of the simulator and describe the methods
used to include lensing effects and observational noise. In
Sect.~\ref{sect:numtest}, we use a galaxy cluster obtained from N-body
simulations and use our code to create mock optical observations of
its central regions. We discuss how several strong lensing features
can be extracted from the simulated images and how the properties of
gravitational arcs can be measured. We also quantify how these
properties depend on the assumed source models. Finally, in
Sect~\ref{sect:conclu} we report our conclusions.

\section{The simulator}
\label{sect:simulator}

In this Section we describe the features included in our simulator for
optical observations. In the following we will assume that:
\begin{itemize}
\item the lenses that are eventually placed in front of the sources
  are obtained from N-body and hydrodynamical simulations. However,
  any deflector can be used for testing the codes, for example
  mass distributions that can be generated using analytic models;

\item a method for calculating the deflection angle fields out of
  the simulated mass distributions has been separately
  implemented. 
\end{itemize}
  
  Our simulator is coupled to an existing ray-tracing
  code that has been widely used in the past \cite[see
  e.g.][]{ME00.1,ME01.1,ME03.2,ME04.1} and that will be described in
  Sect.~\ref{sect:raytr}. We have also implemented other routines for
  calculating the deflection angles from a distribution of particles
  using a tree algorithm \cite[similar to that described in][]{AU07.1}
  and Fast-Fourier-transform methods \citep{BA04.1}. Ray-tracing
  through multiple planes for mimicking the effects of the
  large-scale-structure of the universe has been also implemented
  \citep{PA07.1}. The choice of the implementation is not important
  for this paper, while it will become more relevant in future
  applications of this code. In general, for a ray-tracing simulation
  through a single plane of matter, given a deflection angle map
  $\vec{\hat\alpha}$, the angular position $\vec y$ where a light ray
  is emitted can be easily calculated from the apparent position $\vec
  x$ through the {\em lens equation}:
  \begin{equation}
	\vec  y=\vec x-\frac{D_{\rm ls}}{D_{\rm s}}\vec{\hat\alpha}
        (\vec x) \;,
  \label{eq:leq}
  \end{equation}
where all angles are referred to an arbitrarily chosen optical axis
passing through the observer. The angular diameter distances between
the lens and the source, $D_{\rm ls}$, and between the observer and
the source, $D_{\rm s}$, have been used to re-scale the deflection
angles to the proper lens-source configuration.

Assuming we know the distribution of matter along the line of sight
and to be able to characterize its deflection field, we proceed now to
generate a population of source galaxies, to assign them a given
morphology and surface brightness distribution, to distort them
according to the lensing effect produced by the matter between the
observer and the sources and, finally, to visualize the results of a
virtual observation including all relevant sources of noise and the
convolution with a particular instrument.

\subsection{Galaxy positions, luminosities and redshifts}
\label{sect:gplz}
The virtual observation comprises a user-specified field-of-view. This
defines the size of the light-cone along which source galaxies are
distributed. When generating galaxy positions, we randomly choose the
projected position of the sources on the sky. Their orientation is
also randomly selected. Thus, we do not consider, at the moment,
clustering of galaxies and intrinsic alignments that could represent a
potential source of errors in weak lensing measurements. We will
address this issue in future work.

The source distances, morphological types and intrinsic magnitudes are
generated using observed luminosity functions per given redshift
bin. We choose to adopt the luminosity functions derived from the
VIMOS VLT Deep Survey (VVDS, \cite{LF05.1}), for which different versions exist for
four principal spectral types of galaxies up to $z=1.5$.

\begin{figure}[t!]
  \includegraphics[width=1.0\hsize]{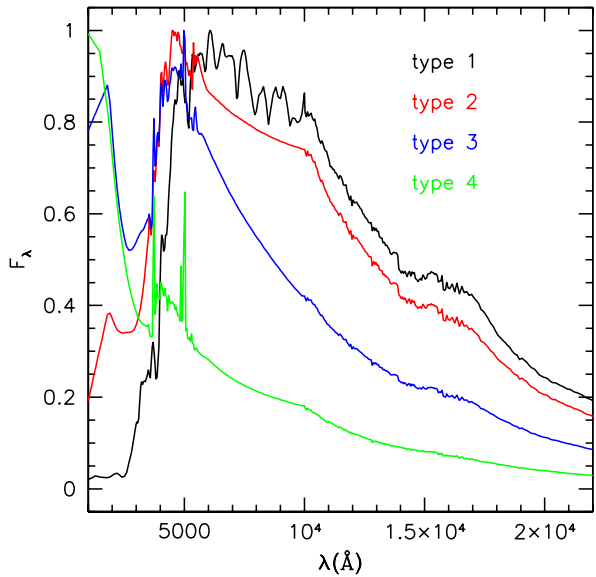}
  \caption{Spectral-energy distributions used to characterize the
    galaxies of different spectral type. The galaxies are divided into
    four principal classes: ellipticals/S0 (type 1), early spirals
    (type 2), late spirals (type 3) and irregulars (type 4). The
    wavelengths are given in $\AA$, while flux is arbitrary.}
\label{fig:seds}
\end{figure}

\cite{ZU06.1} have divided the galaxies in four types, corresponding
to E/S0 (type 1), early spiral (type 2), late spiral (type 3) and
irregular template (type 4). This work has been done by matching the
UBVRI magnitudes with empirical sets of spectral-energy-distributions
(SEDs) \citep{AR99.1}. These SEDs are shown in Fig.~\ref{fig:seds} in
different colors and have been used to characterize the flux of the
synthetic galaxies in our simulations.

The luminosity function for each spectral type and redshift bin is of the form
\begin{eqnarray}
  \phi(M)dM& = &0.4 \phi^{\star}\ln{10}\left(10^{0.4(M^\star-M)} \right)^{\alpha+1} \cdot \nonumber \\ 
  & & \cdot \exp{\left(-10^{0.4(M^\star-M)}\right)}dM \;.
\end{eqnarray}   
It gives the number of galaxies per unit magnitude and per cubic Mpc
\citep{SC76.1}. The characteristic absolute magnitude $M^\star$, the
characteristic density $\phi^{\star}$ and the slope $\alpha$ are
determined by fitting the observed magnitude distributions of galaxies
with the same spectral type and with spectroscopic redshift in the
interval between $z_{\rm min}$ and $z_{\rm max}$. We use the values
reported in Table 3 of \cite{ZU06.1} for the rest frame B-band
(Johnson B filter). The galaxies used in the simulations are generated
such as to reproduce the observed luminosity functions in this
reference band. The magnitudes in the other bands are easily
calculated by using the previous SEDs, suitably normalized such as to
assign the correct luminosity in the B-band, and by convolving them
with the appropriate filter curves.

The luminosity functions for different morphological types exist for
galaxies up to $z=1.5$. For higher redshifts, we adopt the luminosity
functions published by \cite{IL05.1} and by \cite{PAL07.1} and we
extrapolate the fractions of ellipticals, spirals and irregulars from
those observed at lower redshift. In particular at $z \gtrsim 2$ we
assume that only $5\%$ of the galaxies are ellipticals or S0, while
the vast majority of them are spirals and irregulars.

Certainly, important assumptions are made in the process of generating
the population of galaxies filling the light cones. First, the VVDS is
limited to $I$-magnitude 24 (AB magnitude). Thus, the faint tail of
the distribution of galaxy luminosities is extrapolated from a fit
performed on brighter galaxies. Second, the absolute magnitudes in
bands different from the reference B-band are calculated by adopting a
limited number of SEDs. Third, as explained earlier, we can use
luminosity functions per different spectral types only at $z<1.5$. For
larger redshifts, we must use a single luminosity function and
extrapolate from lower redshift the fractions of galaxies with a given
SED.

\begin{figure*}[t!]
  \includegraphics[width=0.5\hsize]{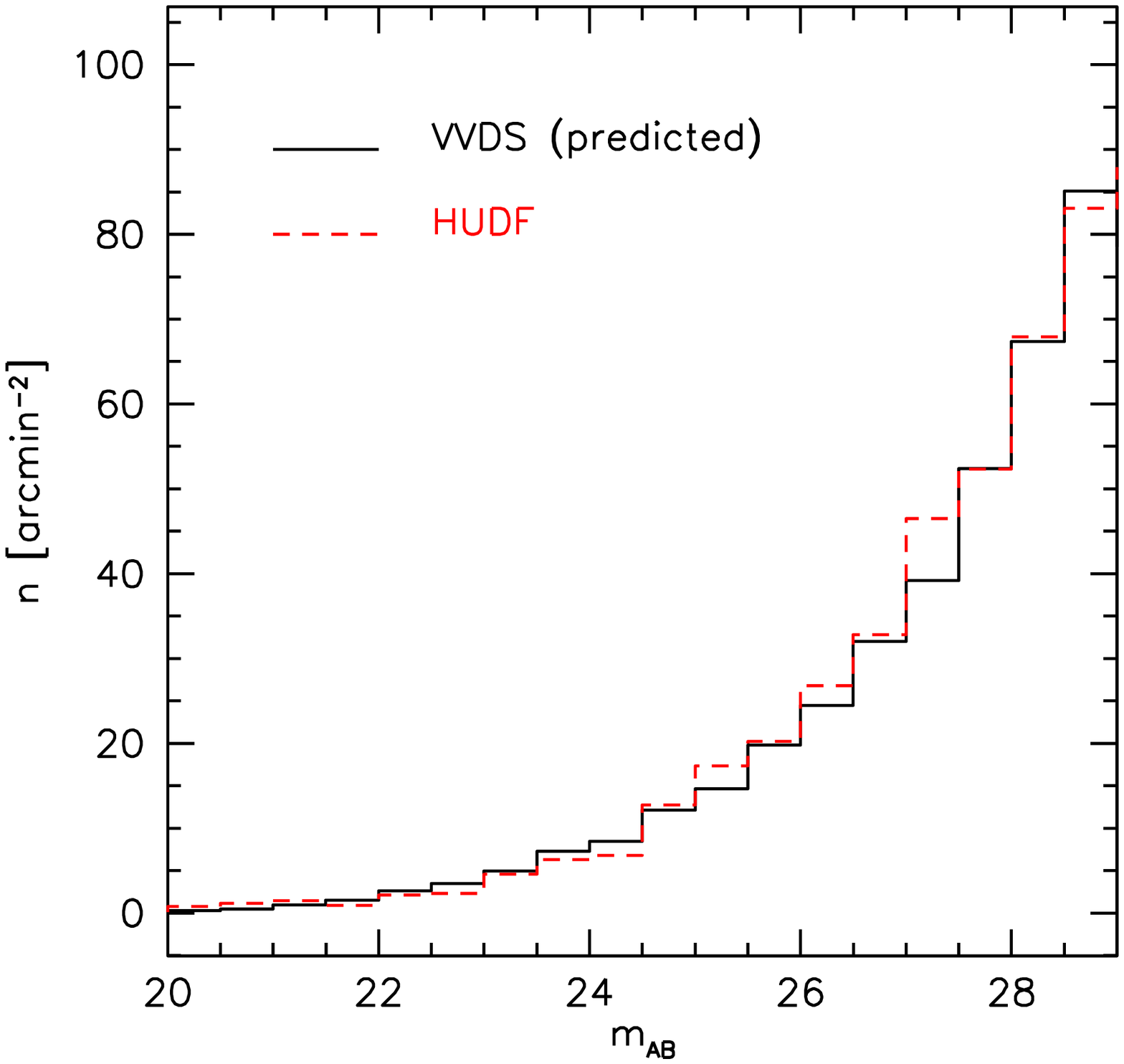}
  \includegraphics[width=0.5\hsize]{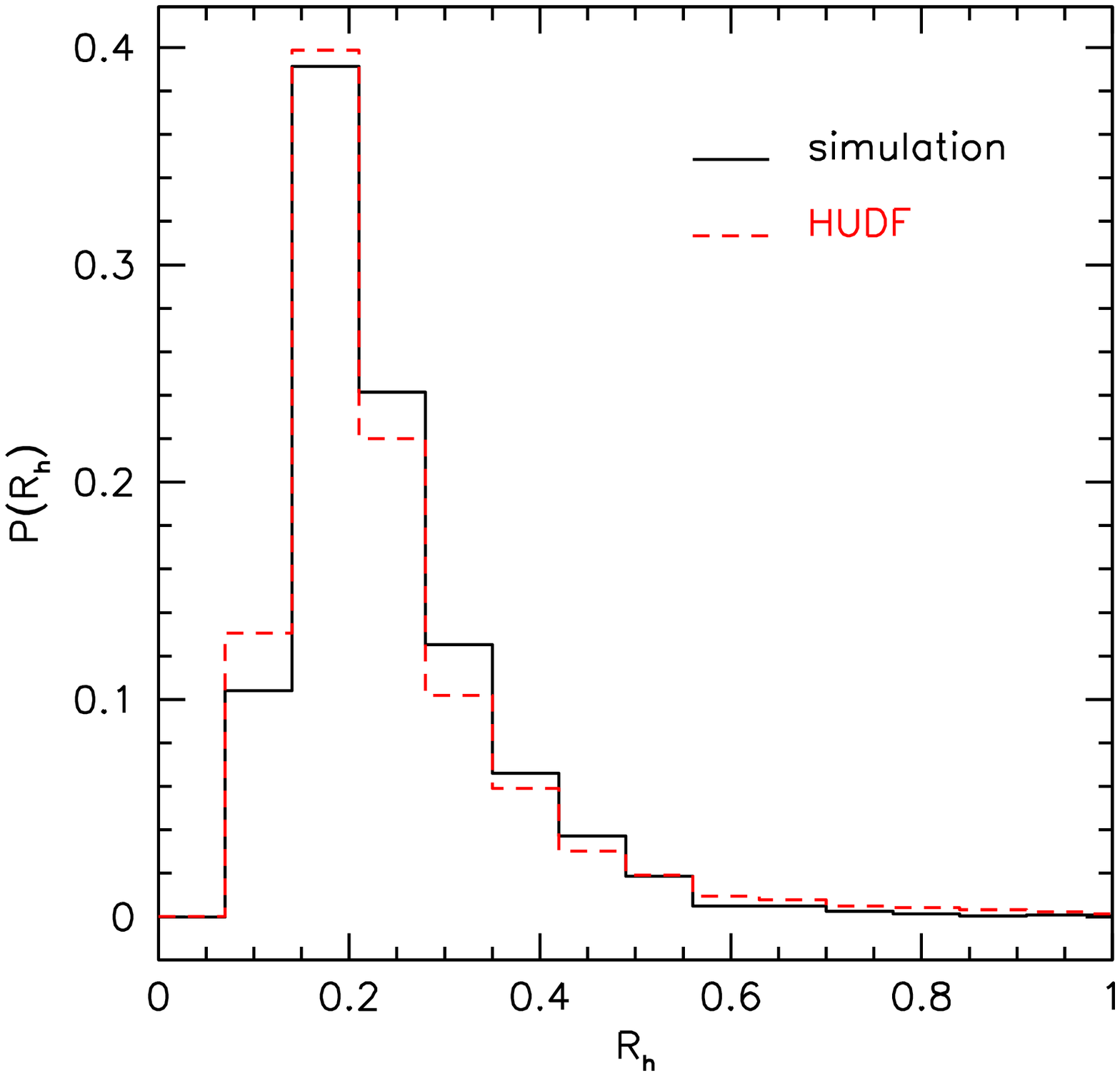}	
  \caption{Left panel: Number counts per apparent magnitude bin and squared arcmin predicted by our method (solid line) and observed in the HUDF
    (dotted line). Magnitudes are given in the $z$-band. Right panel: probability distributions of the galaxy half-light radii in the HUDF (dashed histogram) and in a field simulated with Skylens (solid histogram).}
\label{fig:nz}
\end{figure*}

Although it is natural to expect a dominant fraction of irregulars and
spirals at high redshift, our procedure may lead to significant errors
especially when simulating deep observations. In order to check and
validate our methods, we make the following test. We generate a
population of galaxies as described above and compare the number
counts per magnitude bin and squared arcmin to those measured in the deep
observations like the HUDF. We
find a good agreement between the number counts independent of
the band selected for the experiment. For example, we show in the left panel of 
Fig.~\ref{fig:nz} the comparison between our prediction based on the
VVDS luminosity functions (solid line) and the HUDF (dotted line) in
the $z$-band (F850LP filter on board the Hubble telescope). Although
the selected band is much redder than the reference band and the size
of the HUDF is rather small, the curves are almost perfectly
superposed, confirming the reliability of our results despite the
assumptions made.

\subsection{Galaxy surface-brightness distributions}
The light emission from galaxies falling into the telescope field of
view is modelled using shapelet functions. This approach has been used by \cite{MAS04.1}, although using a different shapelet
decomposition method. According to \cite{RE03.1}, the galaxy
two-dimensional surface-brightness distribution, $I(\vec x)$, can be
approximated by a summation over a set of basic functions,
\begin{equation}
  I(\vec x)=\sum_{n_1,n_2=0}^{\infty} I_{\vec n} B_{\vec n}(\vec x -\vec x_c;
  \beta) \;,
  \label{eq:shapedec}
\end{equation}
where $\vec x_c$ indicates the galaxy centroid, $\vec x=(x_1,x_2)$ is a
two-dimensional vector, and $\vec n=(n_1,n_2)$. The basic functions $B_{\vec
n}(\vec x;\beta)$ are called {\em shapelets} and have the form
\begin{equation}
  B_{\vec{n}}(\vec x,\beta) =\beta^{-1}\phi_{n_1}(\beta^{-1}x_1)\phi_{n_2}(\beta^{-1}x_2) \;,
\end{equation} 
where the one-dimensional functions $\phi_n(x)$ are related to the
Gauss-Hermite polynomials, $H_n(x)$,
\begin{equation}
  \phi_n(x)=[2^n\pi^{1/2}n!]^{-1/2}H_n(x)\exp(-x^2/2) \;.
\end{equation}
The parameter $\beta$ defines the scale of the galaxy image. The coefficients
$I_{\vec n}$ properly weight the contribution of each shapelet function to the
surface-brightness of the galaxy and are given by
\begin{equation}
  I_{\vec n}=\int {\rm d}^2 x I(\vec x)B_{\vec n}(\vec x;\beta) \;.
\end{equation} 

Basic photometric information about the galaxy is readily derived from the
shapelet coefficients. For example, the total flux $I$ is given by
\begin{equation}
  \label{eq:flux}
  I=\pi^{1/2}\beta \sum_{n_1,n_2}^{\rm even} 2^{\frac{1}{2}(2-n_1-n_2)}
\left(\begin{array}{c} n_1 \\ n_1/2\end{array}\right)^{1/2}
\left(\begin{array}{c} n_2 \\ n_2/2\end{array}\right)^{1/2}I_{\vec n} \;,
\end{equation}
and the rms radius $r_{I}$, defined as $r_I^2\equiv\int {\rm d}^2xx^2I(\vec
x)/I$, is given by  
\begin{eqnarray}
  r_I^2& =& \pi^{1/2}\beta^3 I^{-1}\sum_{n_1,n_2}^{\rm even}
  2^{\frac{1}{2}(4-n_1-n_2)}(1+n_1+n_2) \nonumber \\
& & \times
\left(\begin{array}{c} n_1 \\ n_1/2\end{array}\right)^{1/2}
\left(\begin{array}{c} n_2 \\ n_2/2\end{array}\right)^{1/2}I_{\vec n} \;.
\end{eqnarray}

\begin{figure*}[t!]
  \includegraphics[width=0.33\hsize]{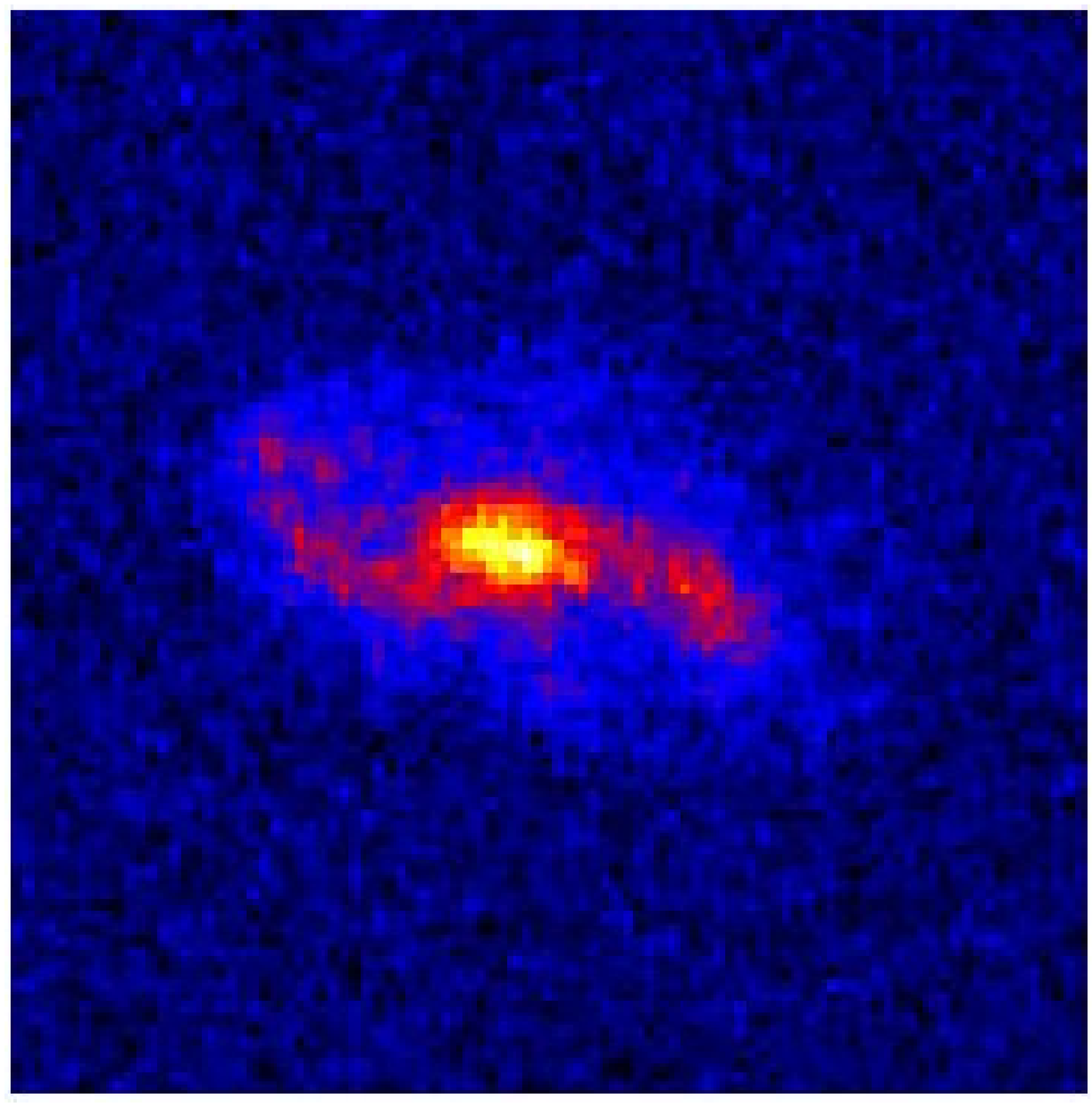}
  \includegraphics[width=0.33\hsize]{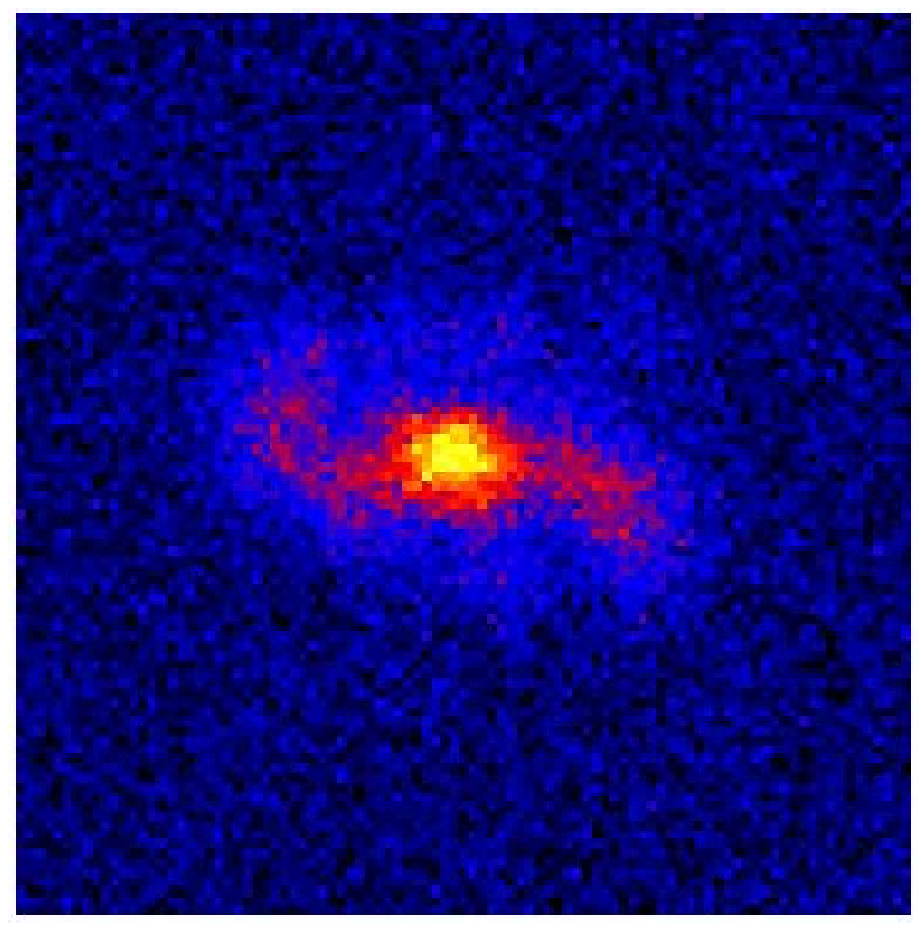}
  \includegraphics[width=0.33\hsize]{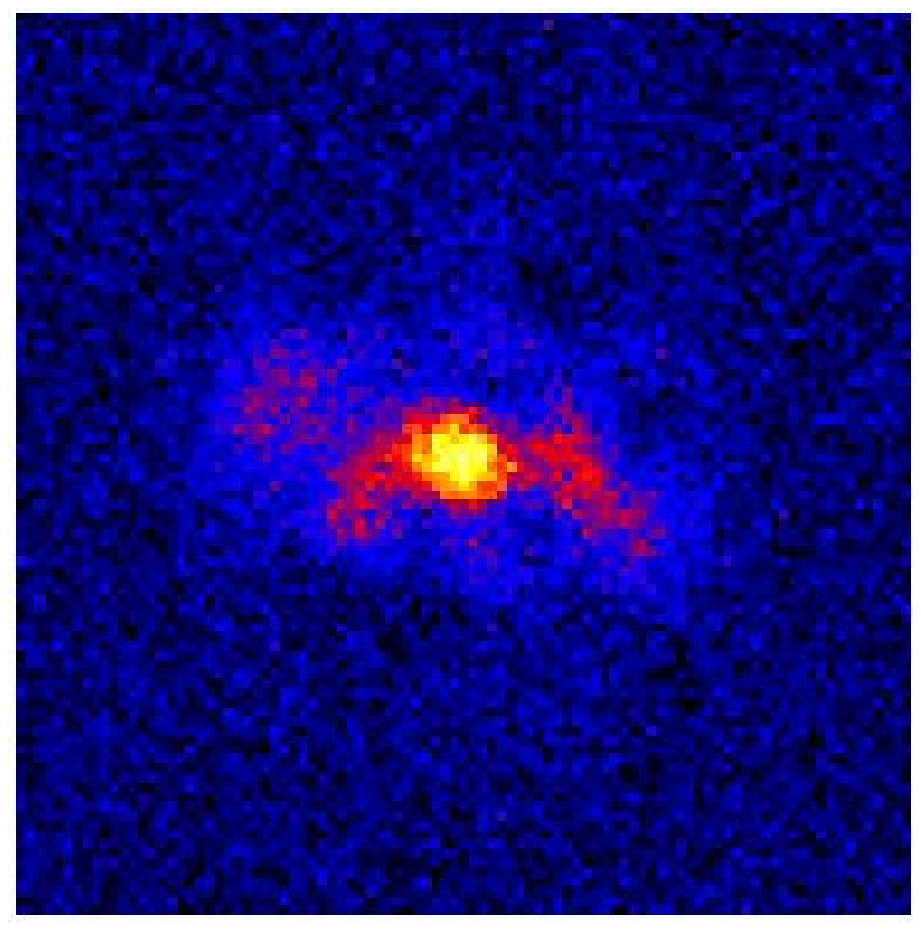}
\caption{Example of generation of synthetic galaxies with shapelets. In the
  left panel the original image of a GOODS galaxy is shown. The galaxy is
  decomposed in $9\times 9$ shapelets and reconstructed in the central
  panel. A second galaxy is then generated from the same shapelet
  decomposition by slightly modifying the coefficients and displayed in the
  right panel. The same color scale is used for all three images. To reproduce
  the background and the noise properties of the image in the left panel, we
  have simulated an observation with HST/ACS (filter F850LP, $t_{\rm
  exp}=10600$ sec). Each frame is $2.9''$ on a side.}
\label{fig:shapeletrec}
\end{figure*}

The shapelet decomposition has proven to be a powerful technique for image
anaysis \citep[see e.g.][]{ML07.1,MAS07.1,GO05.1}. Among the many applications,
\cite{KE04.1} have shown that it can be applied to morphological
classification. In fact, they find that galaxies of different morphological
types separate well in shapelet space. On the other hand, \cite{MAS04.1} showed
that an infinite number of synthetic galaxy images can be easily created from
a set of shapelet coefficients obtained from decomposing real
galaxies. Following these ideas, we aim at using real shapelet decompositions to
characterize galaxy morphologies in our simulator. In detail, we proceed as
follows:
\begin{enumerate}
\item we use the newly developed code by \cite{ML07.1} for
  shapelet-decomposing the images of $\sim 3000$ galaxies from the GOODS
  HST/ACS data \citep{GIA04.1}. For each galaxy, the code provides the best fit set of
  coefficients, $I_{\vec n}^{\rm GOODS}$, together with their errors, and the
  scale parameter, $\beta^{\rm GOODS}$;
\item for about half of the galaxies in the database a spectral
  classification is available, which allows us to
  distinguish between morphological types and to assign a SED to each
  galaxy. For those galaxies which have no classification, we assume
  that they are irregulars or late spirals, since they are typically
  faint and distant blue objects. Based on this classification, the
  galaxy decompositions are divided into four catalogs corresponding
  to the SEDs described in Sect.~\ref{sect:gplz};
\item aiming at generating a galaxy of type M, one entry from the
  corresponding catalog is randomly chosen. Since our catalogs contain a
  limited number of galaxies, which may be smaller than the number of galaxies
  to be generated in the simulation, replications of the same galaxy may
  occur. To avoid that, whenever a galaxy is chosen, we slightly change the
  values of the shapelet coefficients (within their errors) to obtain a new,
  slightly different, galaxy image;
\item the galaxy image is finally rescaled on the basis of the input flux. From the decompositions contained in our shapelet database we build an empirical relation between the scale parameter and the flux. Then, we use this relation and its dispersion to draw the scale parameter $\beta$ that corresponds to the input flux $I$.
Eq.~\ref{eq:flux} however
  states that $I \propto \beta I_{\vec n}$. Thus, to ensure consistency, when
  scaling a GOODS galaxy with scale parameter $\beta_0$ and flux $I_0$ to one with scale parameter $\beta$ and observed flux $I$, we also adjust its shapelet coefficients by multiplying them by $(I/I_0)(\beta_0/\beta)$. In order to verify the reliability of our method, we compared the size distribution of simulated galaxies with that observed in the HUDF. The probability distribution functions of the half-light radii (as outputted by SExtractor, \cite{BE96.1}) in the simulated and in the observed HUDF are shown in the right panel of Fig~\ref{fig:nz}. For the observed HUDF, we refer to the catalog published by \cite{BECK06.1}. The simulated images have been analyzed consistently, i.e. by adapting the SExtractor parameters to those used in the observations. Clearly, the distributions are very similar, although the half-light radii tend to be slightly larger in the simulation (median $R_h=0.211"$) than in the real observation (median $R_h=0.204"$). An even more accurate result will be achieved when our shapelet database is extended using deeper observations, including a larger number of low-flux sources;
\item once a set of coefficients and a scale parameter have been defined, they
  can be used to generate a new galaxy surface-brightness distribution. In
  order to ensure random orientations, we rotate by a random angle
  $\psi \in[0,\pi]$
\end{enumerate}
In Fig.~\ref{fig:shapeletrec} we show an example illustrating the
generation of syntetic galaxies out of a real GOODS galaxy. In the
left panel, we show a spiral galaxy from GOODS. The galaxy is
decomposed into shapelets with maximum order $n_1=n_2=9$. The decomposition is then used
to reconstruct the original galaxy image in the central
panel. Finally, a second galaxy surface brightness distribution is
generated from the same decomposition but after having slightly
changed the values of the shapelet coefficients. As explained above,
the coefficients and the scale parameter may also be rescaled
to change the output flux of the sources, allowing  us to create synthetic
galaxies in arbitrary numbers. Moreover, as it will be shown later,
using shapelets permits us to easily incorporate lensing effects in the
simulations. All images already contain several sources of noise that
will be discussed in the following sections.

\subsection{Observing deep fields}
Once a number of galaxies have been generated, having a luminosity, a SED, a redshift and a shape assigned, we can simulate observations
for a particular instrumental set-up. This implies calculating the number of
photons coming from a patch of the sky which are collected by each element of
the CCD camera, after the light has passed through a sequence of mirrors and
filters, and, eventually, through the Earth's atmosphere.

\subsubsection{Photon counts}
For the implementation of our simulator, we follow the prescriptions
given by \cite{GR04.1} in the construction of the {\em
Large-Binocular-Telescope Camera Image Simulator} (LBCSIM). Given a telescope of diameter
$D$, the number of photons collected by the CCD pixel at $\vec x$ in the
exposure time $t_{\rm exp}$, from a source whose surface brightness is
$I(\vec{x})$ (erg s$^{-1}$cm$^{-2}$Hz$^{-1}$arcsec$^{-2}$), is
\begin{equation}
  n_\gamma(\vec x)=\frac{\pi D^2 t_{\rm exp}p^2}{4 h}I(\vec
  x)\int\frac{T(\lambda)}{\lambda}{\rm d}\lambda \;,
  \label{eq:ngamma}
\end{equation} 
where $p$ is the pixel size in arcsec, $h$ is the Planck constant and
$T(\lambda)$ is the total transmission, which is given by the product of the
atmospheric extinction for a given airmass, $A'(\lambda)=10^{-0.4\cdot{\rm
airmass}\cdot A(\lambda)}$ (airmass$=0$ for observations from space), and the efficiencies of the CCD, $C(\lambda)$, of
the filter, $F(\lambda)$, of the mirror, $M(\lambda)$, and of the optics
$O(\lambda)$,
\begin{equation}
  T(\lambda)=A'(\lambda)\cdot C(\lambda)\cdot F(\lambda) \cdot M(\lambda) \cdot
  O(\lambda) \;.
\end{equation}   
The number of photons is converted into a number of ADUs (Astronomical Data Units) by dividing by the
gain $g$ of the CCD:
\begin{equation}
  ADU(\vec x)=\frac{n_\gamma(\vec x)}{g} \;.
  \label{eq:ADUs}
\end{equation}

\subsubsection{Lensing effects}

Lensing by intervening matter along the line of sight is
applied by mapping the CCD pixel coordinates using the lens equation
\ref{eq:leq}. The pixel at $\vec x$ is now connected to the
position $\vec y$ on the plane where the galaxy surface brightness is
assigned. Then, the surface brightess at $\vec y$ is converted into ADUs using
Eqs.~\ref{eq:ngamma} and \ref{eq:ADUs} and the computed value is assigned to
the pixel at position $\vec x$ on the CCD:
\begin{equation}
  ADU_{\rm lensed}(\vec x)=ADU[\vec y(\vec x)] \;,
\label{eq:ADUmap}
\end{equation}
Note that in the strong lensing regime one coordinate $\vec
y$ can correspond to more than one pixel coordinate $\vec x$.

The application of Eq.~\ref{eq:ADUmap} in the case of arbitrary pixel scales
involves two steps. First, since the deflection angles may be determined on a
grid whose point positions generally differ from those of CCD pixels, the
deflection angle map must be interpolated at each pixel position, $\vec x$. We
use a bi-cubic interpolation scheme, which works well provided the spatial
resolution of the virtual CCD is not too large compared to that of the input
deflection angle map. Second, since the galaxies are distributed in redshift, the deflection angles
must be rescaled by the factor $D_{\rm ls}/D_{\rm s}$ appearing in
Eq.~\ref{eq:leq} for each source. Especially when simulating observations
of large fields, this operation may become time consuming because of the large
number of pixels and sources. To shorten the calculations, we bin the sources
in redshift and project all the galaxies in a given redshift bin on a single
plane. For example, we assign to all galaxies at a redshift between $z_{\min}$
and $z_{\rm max}$ the redshift $z_{\rm s}=(z_{\min}+z_{\rm max})/2$. By
changing the size of the redshift bins, one can increase or decrease the
number of source planes. All sources falling into the same redshift bin can
then be processed at the same time, saving a considerable amount of
computational time.

Note that using the shapelet formalism for representing the source galaxies allows to apply weak lensing transformations through relatively simple operators acting on the shapelet states. For first order lensing, \cite{RE03.1} shows that the transformation can be written as 
\begin{equation}
	ADU_{\rm lensed}(\vec x)\simeq \left[ 1+\kappa(\vec x) \hat{K}+\gamma_i(\vec x)\hat{S}_i\right] ADU(\vec x) \;,
	\label{eq:shapelens}
\end{equation}
where $\kappa$ and $\vec\gamma=(\gamma_1,\gamma_2)$ are the
convergence and the shear, respectively. The operators $\hat K$ and
$\hat S$ can be expressed in terms of the raising and lowering
operators for the quantum harmonic oscillator as
\begin{eqnarray}
  \hat K & \equiv & 1+\frac{1}{2}\left[\hat a_1^{\dag 2} + \hat a_2^{\dag 2}-\hat a_1^{2} - \hat a_2^{2} \right] \\
  \hat S_1 & \equiv & \frac{1}{2}\left[\hat a_1^{\dag 2} - \hat a_2^{\dag 2}-\hat a_1^{2} + \hat a_2^{2} \right] \\
  \hat S_2 & \equiv & \hat a_1^{\dag 2}\hat a_2^{\dag 2}-\hat a_1^{2}\hat a_2^{2}
\label{eq:qop}
\end{eqnarray}
Thus, by using these formulas in combination with
Eqs.~\ref{eq:shapedec}, \ref{eq:ngamma} and \ref{eq:ADUs}, the lensing
distortion can be expressed in terms of transformations of the
shapelet coefficients, $I_{\vec n}$.

\subsubsection{Convolution with the PSF and seeing}
Several sources of noise can then be added to the image to resemble
real observational conditions.

First, the images can be convolved with the instrumental point-spread-function (PSF):
\begin{equation}
  I'(\vec x)=\int {\rm d}^2x' I(\vec x')F(\vec x-\vec x') \;. 
  \label{eq:psf}	
\end{equation}
The convolution can be done in two ways. In analogy to the galaxy
images, a given PSF model, $F(\vec x)$, can be decomposed into
shapelets. Convolution is entirely analytically possible in shapelet
space, given the invariance of the shapelet functions under Fourier
transformations. The full formalism is explained in
\cite{RE03.1}. This approach has been used by \cite{MAS07.2} for image
simulations. The method is efficient for doing convolutions involving
un-lensed or weakly lensed galaxies, i.e. galaxies with known shapelet
decompositions. For very distorted galaxies, like in the strong
lensing regime, applying this method would require to further
shapelet-decompose the lensed images, whose shapelet coefficients
cannot be easily derived from the un-lensed ones (for example using
Eq.\ref{eq:shapelens}). Thus, this would lead to a loss of
computational efficiency. For this reason, when simulating fields
including strong lensing features, we opt for applying standard FFT
techniques for doing convolution operations.

This approach allows to simulate the shape of the PSF very
realistically. However, the PSF is unique for the whole image,
i.e. local variations of the PSF shape were not mimicked so far.

In weak lensing applications, the PSF needs to be corrected in order
to extract the lensing signal from the source ellipticities. Residuals
in the PSF corrections, due for example to the above mentioned
variations of the PSF shape are one of the major sources of error
\citep[see e.g.][]{RH07.1}.

To take this into account in our simulations, we propose the
following approach to model the PSF. The general idea is that
distortions of the PSF shape can be described as an additional
``lensing'' effect. Anisotropies cause the images of stars to acquire
irregular shapes, being stretched along some particular directions as
if an external shear was applied. Such an external shear can be
characterized by a power spectrum, whose amplitude defines the
intensity of the anisotropies and whose shape characterizes the
spatial scales over which it varies.

Let the shear power spectrum be $P_{\gamma}$. We can link the power
spectrum to a source potential power spectrum using standard
definitions. Since the components of the shear are combinations of the
second derivatives of the lensing potential,
\begin{eqnarray}
   \gamma_1& = & \frac{1}{2}\left(\frac{\partial^2 \psi}{\partial x_1^2}-\frac{\partial^2 \psi}{\partial x_2^2}\right) \\
    \gamma_2&=&\frac{\partial^2 \psi}{\partial x_1 \partial x_2} \;,
\end{eqnarray}
in the Fourier space we have that
\begin{eqnarray}
    \hat\gamma_1& = & \frac{k_1^2-k_2^2}{2} \hat\psi \\
    \hat\gamma_2 & = & k_1k_2\hat\psi \;,
\end{eqnarray}  
where the hat denotes Fourier transforms and $\vec k=(k_1,k_2)$ is the wave vector. The power spectrum of the source potential is thus
\begin{equation}
    P_\psi=\frac{4}{k^4}P_\gamma \;.
\end{equation}  

Using this formalism, the distortions of the PSF can be introduced by applying additional lensing to the images already convolved with a PSF function. Given a potential $\psi$, the corresponding deflection angle field is
\begin{equation}
  \vec\alpha=\vec\nabla \psi \;,
\end{equation} 
and in Fourier space
\begin{eqnarray}
  \alpha_1 & = & -\mathrm{i}k_1\hat\psi \\
  \alpha_2 & = & -\mathrm{i}k_2\hat\psi \;.
\end{eqnarray}
Thus, the power spectra of the two components of the deflection angle are:
\begin{eqnarray}
	P_{\alpha_1}&=&k_1^2P_{\psi}=\frac{k_1^2}{k^4}P_\gamma \\
	P_{\alpha_2}&=&k_2^2P_{\psi}=\frac{k_2^2}{k^4}P_\gamma \;.
\end{eqnarray}

For simplicity, we assume that $P_{\gamma}$ is a Gaussian,
\begin{equation}
   P_\gamma(k)=A_\gamma \exp{-\frac{k^2}{k_{\rm cut}^2}} \;,
\end{equation}
whose amplitude is $A_\gamma$ and whose characteristic scale is $k_{\rm cut}$. This last parameter sets the scales over which spatial variations of the PSF shape occur.

Thus, we use $P_\gamma$ to generate maps of the deflection angle components. For doing that we assume that these maps are Gaussian, with zero mean and variance defined by the amplitude of the input power spectrum. The deflection angles are used to lens the images as explained in the previous section.

When simulating ground based observations, seeing is simulated by further
convolving the image with a Gaussian $G$ of size $\sigma$:
\begin{equation}
  I_\sigma(\vec x)=\int {\rm d}^2x' I'(\vec x')G(\vec x-\vec x') \;. 
  \label{eq:seeing}
\end{equation}

\begin{figure}[t]
  \includegraphics[width=\hsize]{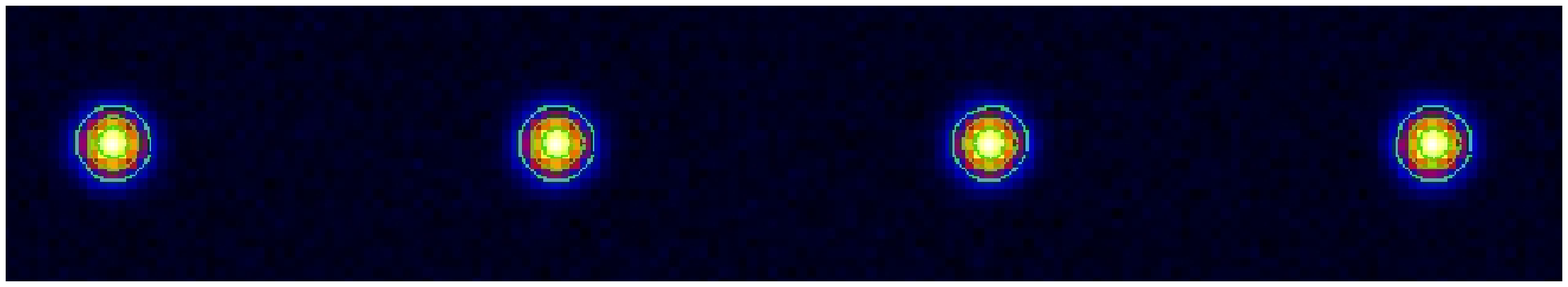} \\
  
  \vspace{0.2mm}
  \includegraphics[width=\hsize]{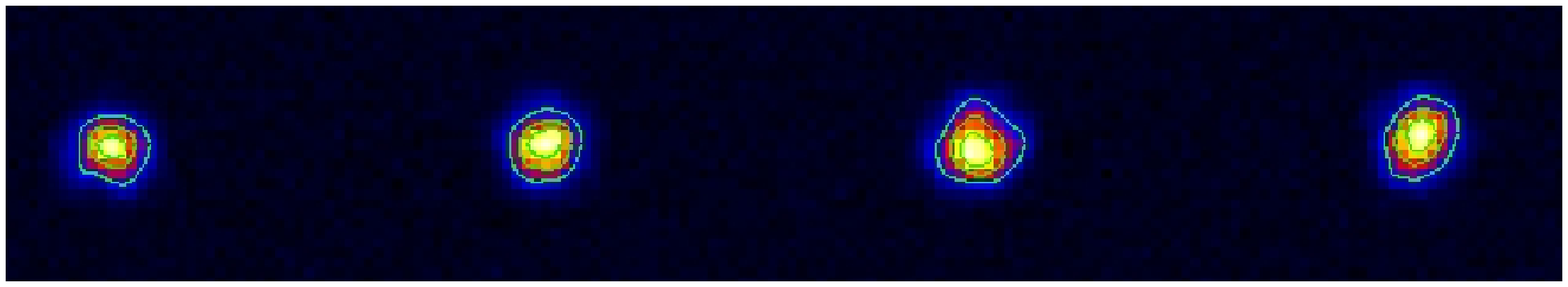} \\
  
  \vspace{0.2mm}
  \includegraphics[width=\hsize]{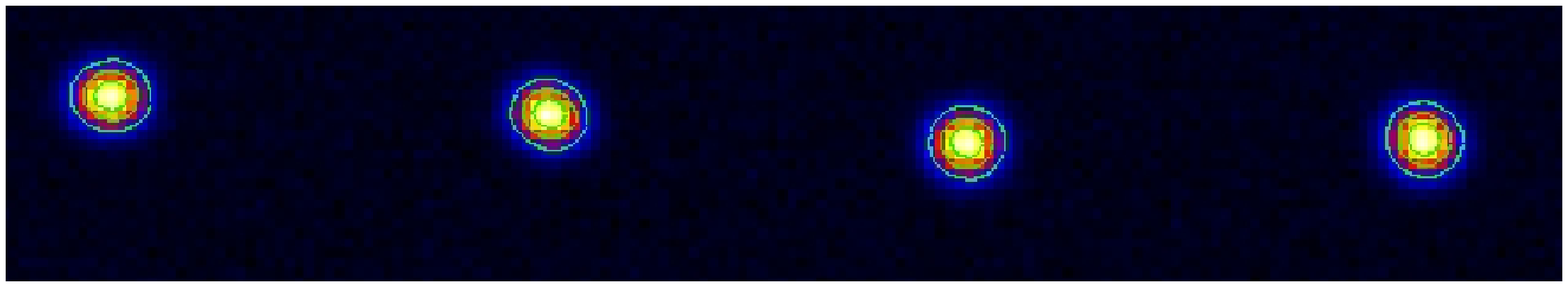}
  \caption{Images of an array of stars observed with an LBT-like
    telescope. The angular separation between the stars is $10"$ along
    the $x$ axis. In the top panel, an isotropic PSF (FWHM=$0.6"$) is
    mimicked. In the middle panel, anisotropies varying on scales of
    $\sim 1"$ are included. In the bottom panel we simulate
    larger-scale variations of the PSF shape on scales of $\sim 15"$.}
\label{fig:psf}
\end{figure}

In Fig.~\ref{fig:psf} we show some examples. In the top panel, an
array of stars convolved with an isotropic PSF with Gaussian profile
and FWHM of $0.6"$ (including seeing) is shown. The angular separation
of the stars along the $x$ axis is $10"$. In the middle panel,
we introduce variations of the PSF shape on scales of $\sim 1"$. The
amplitude of the distortions is intentionally quite large and
correspond to $A_{\gamma}=0.9$. Instead, in the bottom panel we mimic a
PSF that slowly varies on scales of $\sim 15"$. These examples show
that by properly choosing the parameters $A_\gamma$ and $k_{\rm cut}$
a suitable level of noise in the PSF can be added.

 
\subsubsection{Sky background and photon noise}
The number of ADUs per pixel from the sky background are given by
\begin{equation}
  ADU_{\rm sky}=\frac{\pi D^2 t_{\rm exp}p^2}{4 h
  g}\int\frac{T(\lambda)S(\lambda)}{\lambda}{\rm d}\lambda \;,
\end{equation}
where $S(\lambda)$ is the sky flux per square arcsec at a given
wavelength and $T(\lambda)$ is calculated now at ${\rm
  airmass=0}$. The sky flux is modeled differently for observations
from the ground and from space. For ground based observations, we use
different night sky spectra according to the different phases of the
moon. The spectra are those used in \cite{GR04.1}. They correspond to
0, 3, 7, 10 and 14 days after the dark moon. For simulating the sky as
seen from the space, we adopt several spectra of zodiacal light,
representing the typical variations of the zodiacal background
radiation and of the earthshine \citep{GIA02.1}.

Photon noise is assumed to be Poissonian. For each pixel the
noise (in ADUs) is generated by adding random deviates with Gaussian
probability distribution of mean zero and width
\begin{eqnarray}
  \sigma_N(\vec x)&=&\left\{n_{\rm
  exp}\left(\frac{RON}{g}\right)^2+\frac{ADU(\vec x)+ADU_{\rm
  sky}}{g}\right. \nonumber \\
  & & +\left. \left(f+\frac{a^2}{n_{\rm exp}^2}\right)[ADU(\vec x)+ADU_{\rm
  sky}]^2\right\}^{1/2} \;.
\end{eqnarray}
In the previous formula $RON$ is the read-out noise of the chip, $n_{\rm exp}$
is the number of exposures, $a$ is the flat-field term, which we fix at
$a=0.005$ following \cite{GR04.1}. The term $f$ indicates the flat-field
accuracy, which is determined by the number of flat-field exposures and by the
level of the sky background as
\begin{equation}
  f=(N_{ff}\cdot B\cdot g)^{-1} \;.
\end{equation}
A more detailed explanation of these formulas is given in Section 4.2 of
\cite{GR04.1}. 

\subsubsection{Producing the images}

\begin{figure}[t!]
  \includegraphics[width=0.49\hsize]{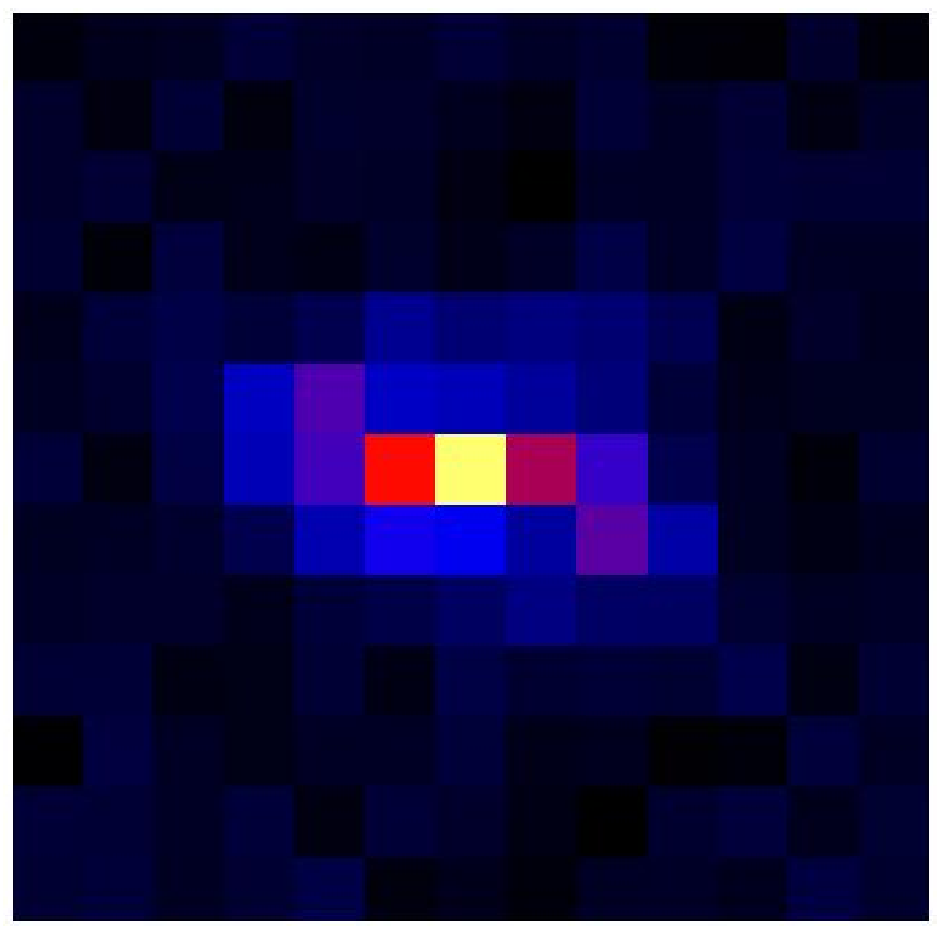}
  \includegraphics[width=0.49\hsize]{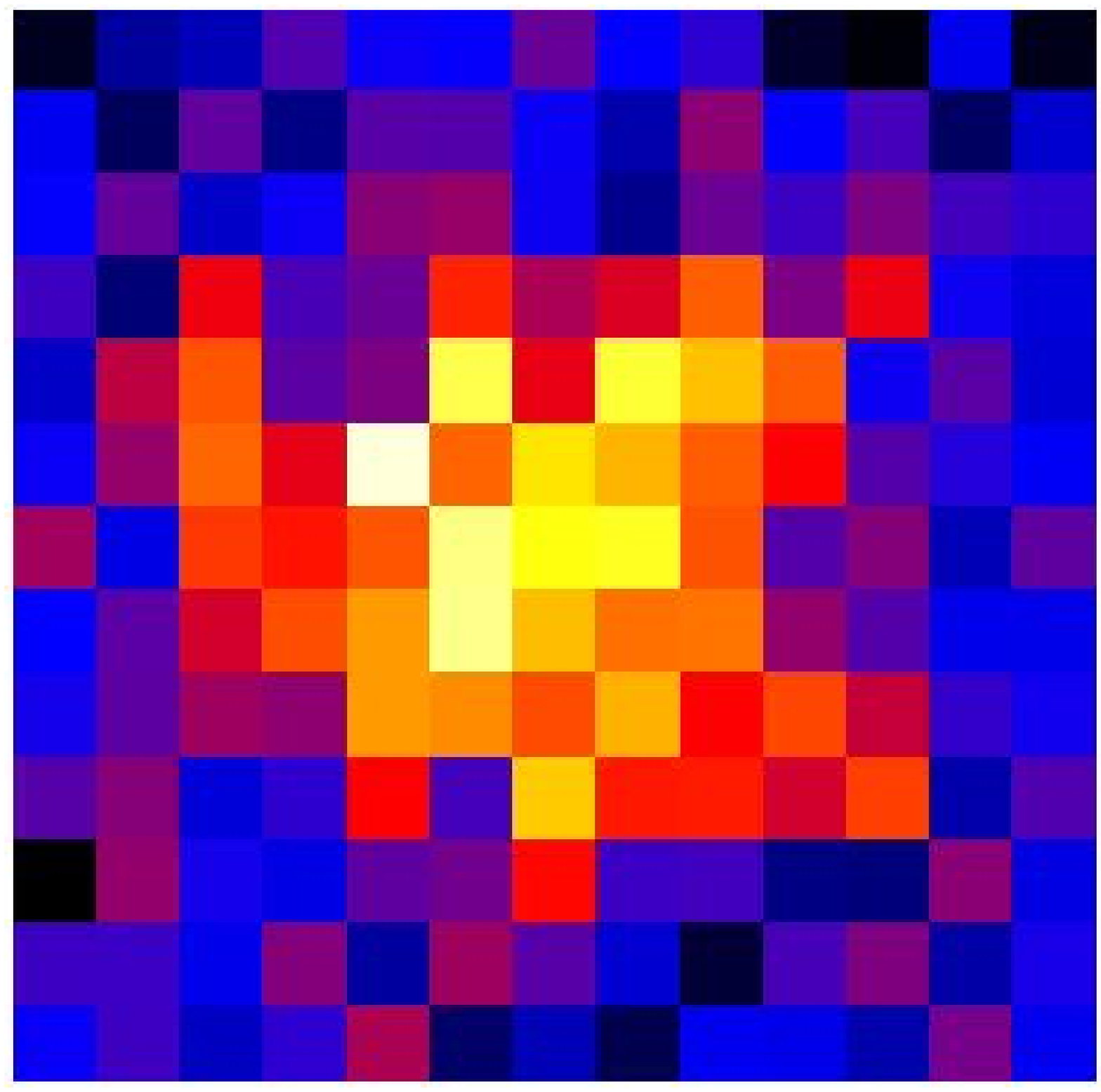}
\caption{The same galaxy displayed in the central panel of
  Fig.~\ref{fig:shapeletrec} is shown here as it should be observed in the
  $I$-band by LBT with a diffraction limited PSF (left panel) and in bad seeing
  conditions ($\sigma=1.2''$, right panel). The exposure time is
  $t_{\exp}=2000$ sec and the airmass is 1. The scale of each frame is
  $2.9''$.}
\label{fig:shapeletground}
\end{figure}

By properly setting the total efficiency (using the correct efficiency curves
for optics, mirrors and CCDs) observations from several instruments can be
simulated. As an example, we show in Fig.~\ref{fig:shapeletground} how the
same galaxy displayed in the central panel of Fig.~\ref{fig:shapeletrec}
would appear when observed with the LBT with a diffraction limit PSF (left panel) and in bad seeing conditions ($\sigma=1.2''$; right panel). The exposure time is $2000$
sec with ${\rm airmass}=1$. The pixel size of the LBT camera is $0.224''$, in
contrast to the $0.03''$ of HST/ACS (after drizzling). The spiral structure of
the galaxy is thus not resolved by LBT. Seeing blurs the image, destroying
most of the remaining morphological information.

\begin{figure}[t]
  \includegraphics[width=\hsize]{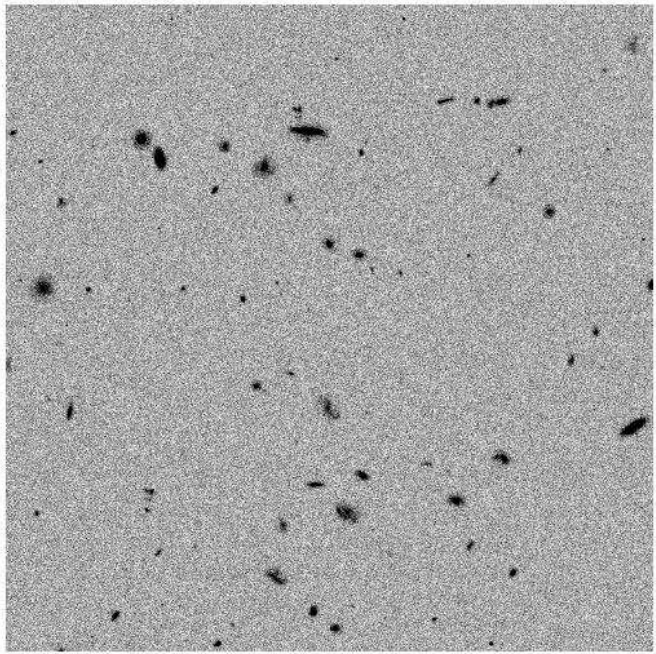}
\caption{Simulation of a deep field observed by HST/ACS in the $z$-band. The
  exposure time is $t_{\rm exp}=10600$ sec and the field of view is $1'\times
  1'$.}
\label{fig:deepfield}
\end{figure}

Combining several galaxies into the same field, we are able to produce
artificial galaxy fields. One example is given in Fig.~\ref{fig:deepfield}, where we
simulate a deep field of $1'\times 1'$ observed with the HST/ACS with an exposure
time of $10600$ sec. In this simulation, no lensing effects have been included. As demonstrated here, using our
procedure for generating artificial galaxies we can produce very realistic
simulated images of the sky. 

\subsubsection{Foregrounds}
\label{sect:frg}

The light emission from foreground galaxies can be included in simulated images
adopting the same procedure employed for background sources. We note that
cluster galaxies in particular represent a potential problem for lensing
analyses. For weak lensing studies, misidentification of cluster members as
background sources can bias the shear measurements. In addition the bright
galaxies concentrated near the cluster centre (which corresponds to the strong
lensing regime) complicate the detection of lensed sources. 

In order to include realistic galaxy clusters in our simulated images,
we have used results from a semi-analytic model of galaxy formation
coupled to a cluster N-body re-simulation. The semi-analytic model used
in this study is described in \cite{DL07.1} and we refer to the
original study (and references therein) for more details about the
physical processes modelled.  The model is used to generate a
catalogue containing positions and luminosities of model galaxies at
different redshifts. This information is used to simulate galaxy
images as described in the previous section.  As in previous work, we
determine the morphology of our model galaxies by using the B--band
bulge--to--disc ratio together with the observational relation by
\cite{SI86.1} between this quantity and the galaxy morphological type.

\begin{figure}[t]
  \includegraphics[width=\hsize]{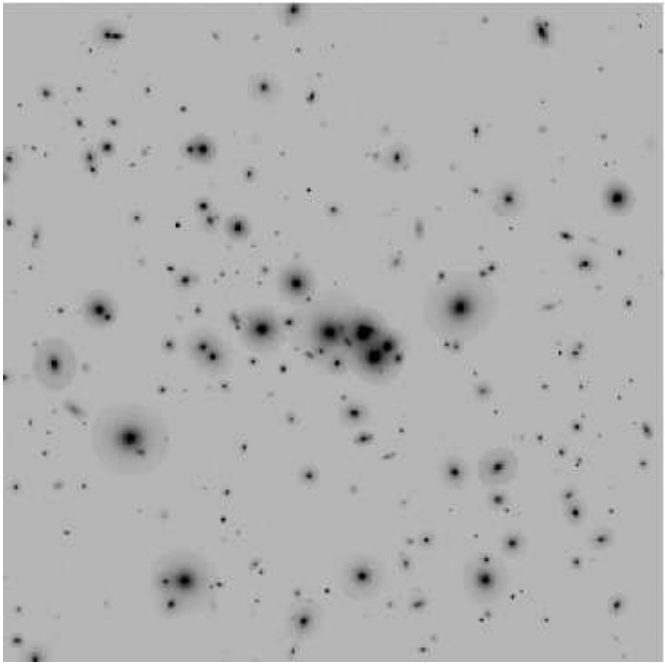}
\caption{Virtual observation in the $z$-band by HST/ACS of a galaxy cluster at
  $z\sim 0.3$. The exposure time is $t_{\rm exp}=1000$ sec and the field of
  view is $100"\times 100"$. The spatial distribution and luminosities of the
  cluster galaxies have been obtained from semi-analytic models. The lensing effects are disabled in this simulation.}
   \label{fig:clusterfield}
\end{figure}

An example of a simulated observation of a galaxy cluster is shown in
Fig.~\ref{fig:clusterfield}. The cluster is at redshift $z\sim 0.3$. It is
observed with HST/ACS in the $z$-band with an exposure time of $1000$ sec.  The
cluster is dominated by some very bright galaxies ($L \sim 10^{12}L_{\odot}$)
in the central part. They are expected to complicate significantly the
observability of lensing features, in particular radial arcs, occurring near
the cluster center.

\section{Applications of the method}
\label{sect:numtest}
In this Section, we discuss one possible application of our simulator, i.e. simulating gravitational arcs in the center of galaxy clusters. Then we apply some techniques, that can be used also in real observations, to measure the properties of the arcs detected in the images.

\subsection{The cluster model}

The cluster used in this paper is one the clusters previously studied by \cite{DO05.1}. Its lensing properties were investigated also by
\cite{PU05.1} and by \cite{ME07.1}. A detailed description of the simulation can be found
in these papers. Its reference name is $g1$. Although several versions
of this cluster exist that include the treatment of the gas component
with several physical processes, we have chosen to use for the present
paper the simplest simulation, where the cluster contains only dark
matter particles.

The halo has a mass of $\sim1.4\times 10^{15}\:h^{-1}M_\odot$ (resolved by more than a million particles within the virial region), thus it
represents a very massive and efficient strong lens at $z=0.3$. We
have chosen this redshift because it is close to where the strong
lensing efficiency of clusters is the largest for sources at $z_{\rm
  s} \gtrsim 1$ \citep{LI05.1}.

Although we are investing the lensing property of this halo only at at
$z=0.3$, during the simulation 92 time slices were saved from redshift
$60$ to $0$.  These are equispaced in time. For each snapshot, we
computed group catalogues and their embedded substructures using a
standard friends-of-friends algorithm with a linking length of 0.2 in
units of the mean particle separation, and a modified version of the algorithm SUBFIND
\citep{SPR01.1} that was extended for future applications to hydro-simulations.  Substructure catalogues were then used to construct
merger history trees as described in \cite{SP05.1} and
\cite{DL07.1}. These merger trees are the basic input needed for the
semi-analytic model described in Sect.~\ref{sect:frg}.




\begin{figure}[t]
  \includegraphics[width=\hsize]{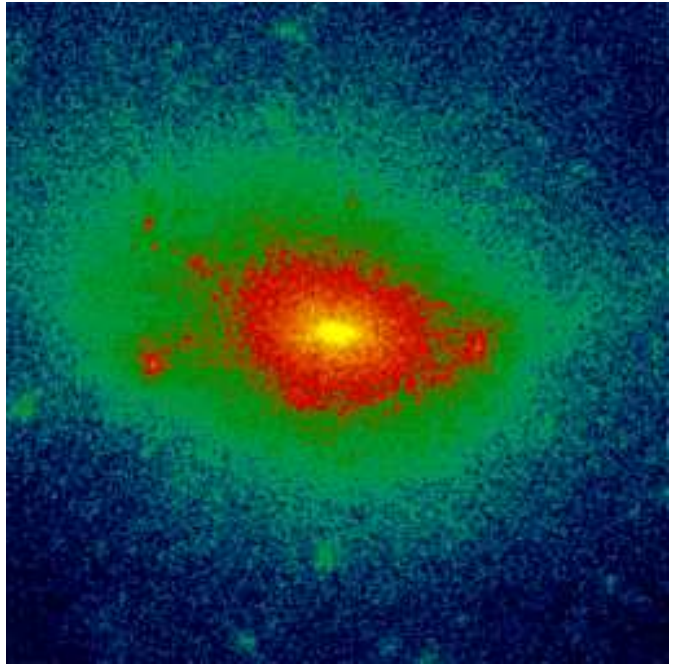}
\caption{Surface density map of cluster $g1$ corresponding to projecting the cluster along the $z$-axis of the simulation box. The side-length of the image is $1.5\,h^{-1}$Mpc comoving (corresponding to $\sim 372"$).}
\label{fig:sden}
\end{figure} 

The surface density map corresponding to one projection of the cluster
is shown in Fig.~\ref{fig:sden}. The size of the image is
$(1.5\,h^{-1}$Mpc)$^2$. The inner region appears quite elliptical and
devoid of large substructures. Secondary massive clumps of matter are
located at distances $\gtrsim 400\,h^{-1}$kpc from the cluster center.

\subsection{Calculations of deflection angles}
\label{sect:raytr}

Ray-tracing simulations are carried out using the technique described in
detail in several earlier papers (e.g.~\citealt{BA98.2,ME00.1}).

We select a cube of $3\,h^{-1}$Mpc comoving side
length, centred on the halo centre and containing the high-density
region of the cluster. The particles in this cube are used for
producing a three-dimensional density field, by interpolating their
position on a grid of $1024^3$ cells using the {\em Triangular Shaped
Cloud} method \citep{HO88.1}. Then, we project the three-dimensional
density field along the coordinate axes, obtaining three surface
density maps $\Sigma_{i,j}$, used as lens planes in the following
lensing simulations. 

The lensing simulations are performed by tracing a bundle of $2048
\times 2048$ light rays through a regular grid, covering the central
sixteenth of the lens plane. This choice is driven by the necessity to
study in detail the central region of the cluster, where critical
curves form, taking into account the contribution from the surrounding
mass distribution to the deflection angle of each ray.

Deflection angles on the ray grid are computed as follows. We first define a grid of $256\times256$
``test'' rays, for each of which the deflection angle is calculated by
directly summing the contributions from all cells on the surface
density map $\Sigma_{i,j}$,
\begin{equation}
  \vec \alpha_{h,k}=\frac{4G}{c^2}\sum_{i,j} \Sigma_{i,j} A
  \frac{\vec x_{h,k}-\vec x_{i,j}}{|\vec x_{h,k}-\vec x_{i,j}|^2}\;,
\end{equation}  
where $A$ is the area of one pixel on the surface density map and
$\vec x_{h,k}$ and $\vec x_{i,j}$ are the positions on the lens plane
of the ``test'' ray ($h,k$) and of the surface density element
($i,j$). We avoid the divergence when the
distance between a light ray and the density grid-point is zero by
shifting the ``test'' ray grid by half-cells in both directions with
respect to the grid on which the surface density is given. We then
determine the deflection angle of each of the $2048\times2048$ light
rays by bi-cubic interpolation between the four nearest test rays.

\subsection{Observations}
For simulating observations of our numerical cluster, we consider an
instrument with the characteristics of the planned space telescope
DUNE ({\em Dark Universe Explorer}; {\tt
  http://www.dune-mission.net}). This mission was recently proposed to
ESA within its "Cosmic Visions" programme. It is targetted at studying
the dark components of the universe with a wide field imager, in
particular through weak lensing. However, thanks to the good spatial
resolution, the panoramic field of view ($0.5$ square degrees) and
high sensitivity, this instrument will be extremely useful for many
other studies including strong lensing, galaxy formation and
evolution, baryonic acoustic oscillations and even planet searches.

The motivation for adopting a future instrument for our virtual
observation is twofold. On one hand, we would like to illustrate that
the simulator is a powerful tool for testing techniques that are
commonly applied to observational data. On the other hand, we would
like to highlight that codes like this can help to define the
scientific tasks that can be fulfilled by future missions.

\begin{figure}[t]
  \includegraphics[width=\hsize]{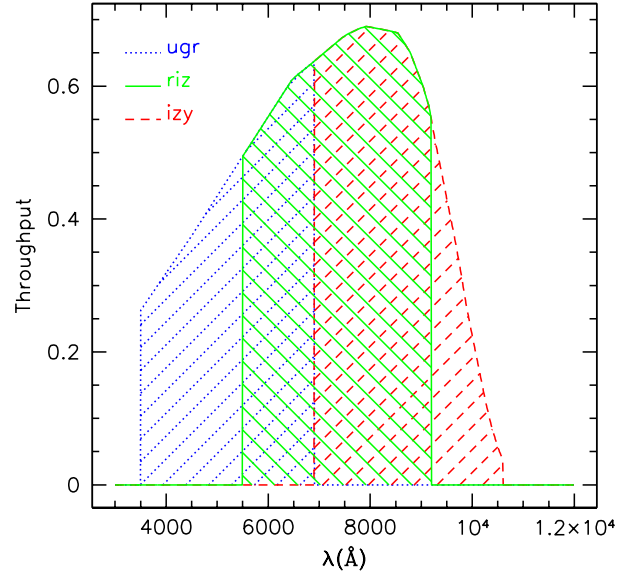}
\caption{Throughputs assumed in carrying out the simulations with the DUNE satellite. Three broad-band filters are used: $ugr$ (dashed lines), $riz$ (solid lines), $izy$ (courtesy of A. Amara, J. Rhodes and the DUNE collaboration members). Note that the figure does not reflect the final choice for the filters to be mounted on DUNE (see text for more details).}
\label{fig:filters}
\end{figure}

DUNE is planned to have a mirror diameter of 1.2 meters, a spatial
resolution of $0.10"$/pixel in the visible and a total efficiency
close to $\sim 70\%$. During the proposal preparation, three
broad-band filters where used, namely a $u+g+r$ ($ugr$), a $r+i+z$
($riz$) and a $i+z+y$ ($izy$) filter. In the proposed version, DUNE will not provide 3 bands in the visible. On the contrary, there will be one filter in the visible ($riz$) and 3 filters in the NIR ($Y,J,H$). The tests that are discussed below have been performed using the $riz$ filter. Just for the purpose of illustrating the capabilities of the simulator for mimicking observations in multiple bands, we used the $ugr$ and the $izy$ filters (see Fig.~\ref{fig:rizobs} below). The throughputs that we assume, including the quantum efficiency of the detector and the optical
design, are shown in Fig.~\ref{fig:filters}. The PSF is expected to
have a FWHM of $0.23"$ with ellipticity smaller than $6\%$.
 
\begin{figure*}[t]
  \includegraphics[width=0.335\hsize]{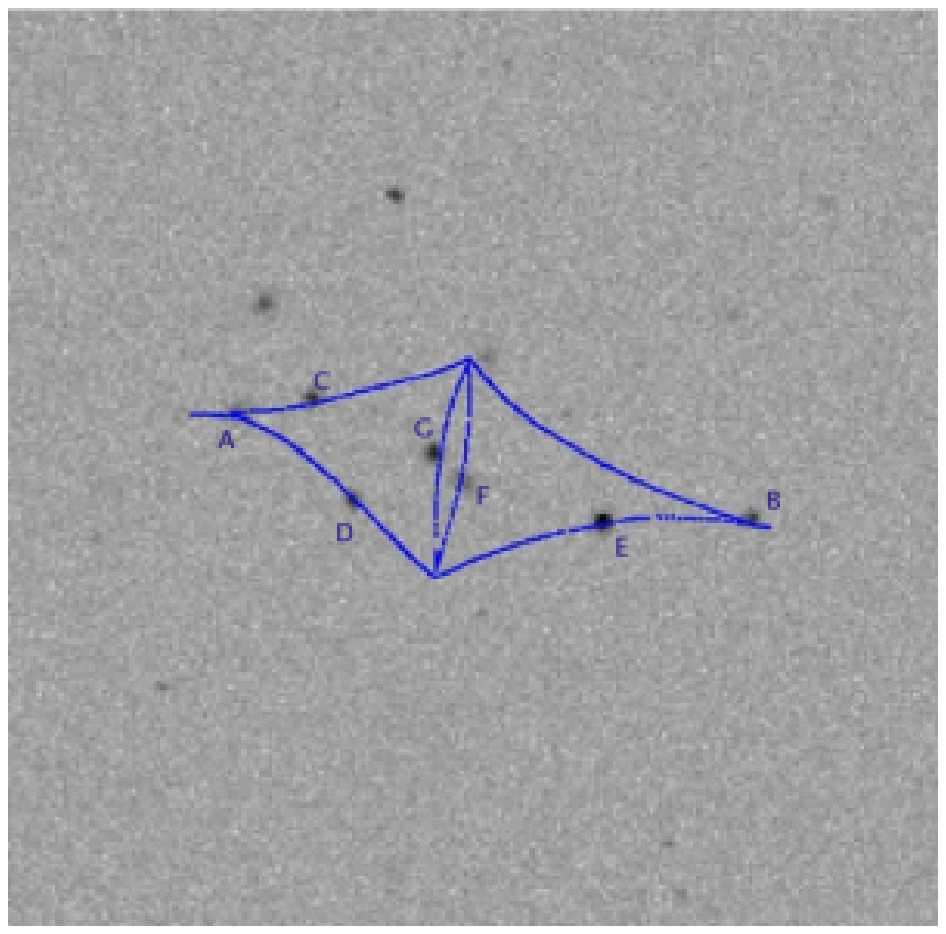}
  \includegraphics[width=0.33\hsize]{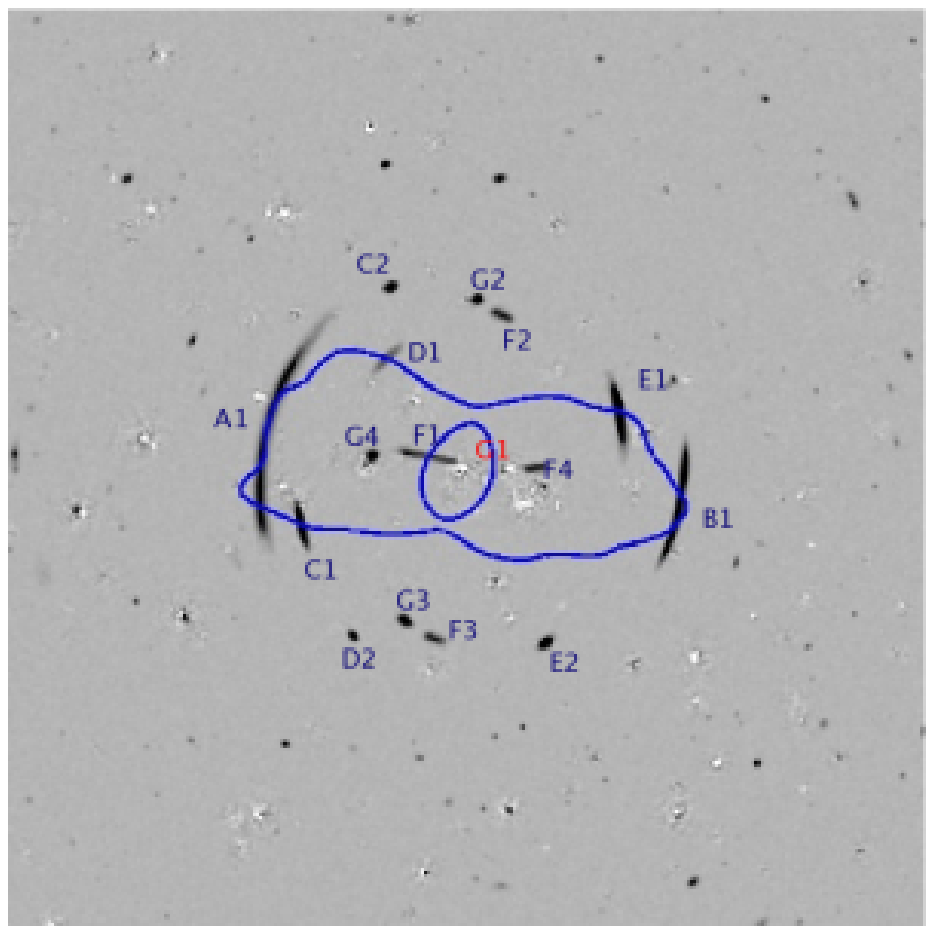}
  \includegraphics[width=0.332\hsize]{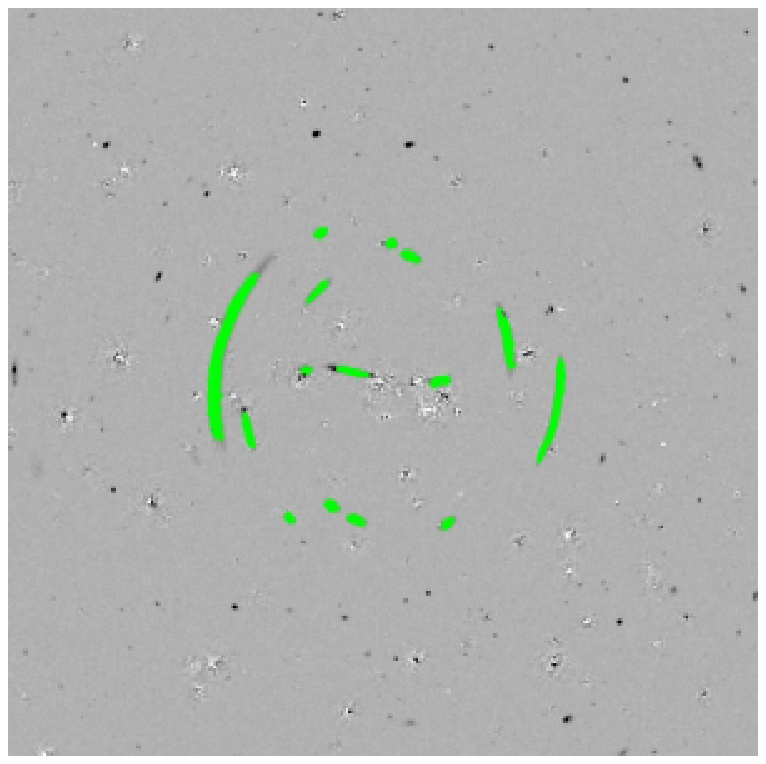}
  \caption{Left panel: positions of the sources relative to the
    caustics of the lens. The image is $25"\times 25"$ wide. Central
    panel: $riz$ image of the cluster g1 after subtracting the cluster
    galaxies using magnitude and size cuts (see text for more
    details). The size of the image is $100"\times 100"$. The images
    of the sources in the left panel are indicated by the letters,
    followed by numbers corresponding to image multiplicity. The image
    G1 is not visible, but we mark in red the position where it should
    be. Right panel: the points belonging to the lensed images are
    marked in green.}
\label{fig:cau_lens}
\end{figure*}

A virtual observation of the inner $100"\times 100"$ region of cluster
$g1$ is shown in Fig.~\ref{fig:rizobs}. It is a composite $ugr$, $riz$
and $izy$ image, corresponding to an exposure time of $1000$ sec. In these tests, we assume an exposure time of $1000$ seconds, although it has been proposed that DUNE exposures will be longer ($1500$ seconds). The galaxies were drawn from our the shapelet library, as explained earlier. In
order to facilitate the formation of strong lensing features, we
intentionally placed seven "test" galaxies close to the caustics of
the lens. They are shown in the left panel of Fig.~\ref{fig:cau_lens},
where we run a simulation without the foreground lens and we zoom over
the region where the caustics are located. The caustics are also shown
in blue. For simplicity, we assume that the "test" sources are all at
$z_s=1.5$.  Two galaxies, labeled A and B, are placed very close to
the cusps of the tangential caustic, thus they are expected to produce
extended cusp arcs. Two other galaxies (F,G) are placed along the
radial caustic and their images will be radially elongated arcs. The
remaining "test" sources (C,D,E) are located along the fold of the
tangential caustic, thus they should be lensed to form fold arcs. The
lensed images are presented in the central panel of
Fig.\ref{fig:cau_lens}.

\begin{figure}[t]
  \includegraphics[width=\hsize]{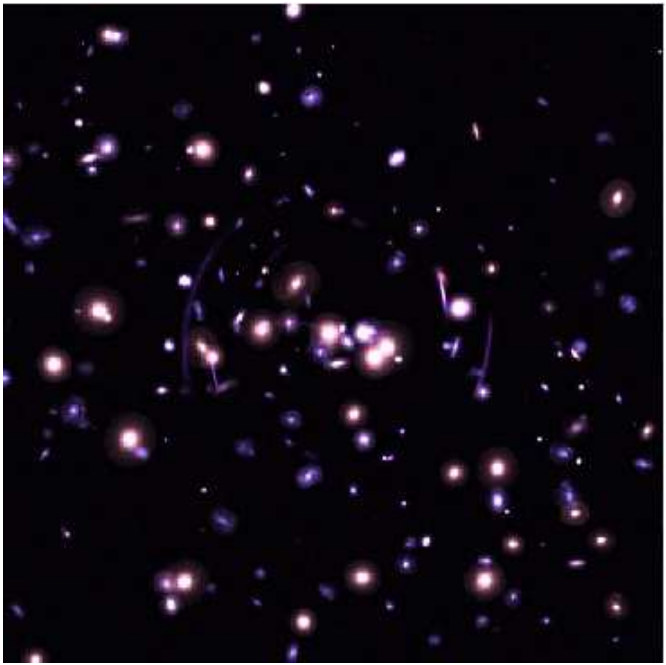}
\caption{Composite $ugr$+$riz$+$izy$ image of the cluster $g1$. Several strong lensing features are present in the image. The FOV size is $100"\times100"$.}
\label{fig:rizobs}
\end{figure} 

Some of the arcs, and especially the radial ones, are partially or
totally hidden by the light of the foreground galaxies, as can be seen
in Fig.\ref{fig:rizobs}. On the basis of the cluster merger tree, the
semi-analytic models predict the presence of a very bright galaxies
near the cluster center. These render accurate measurements of arcs
shapes behind them problematic. In real cases, these galaxies should
be removed \cite[see e.g.][]{SA03.1}. Thus, we subtract the foreground
galaxies from the image by fitting their surface brightness profiles
with Sersic profiles. This can be done with a suitable software like
{\rm GALFIT} \citep{PE02.4}.

We show the image after removing the foreground light in the central
panel of Fig.~\ref{fig:cau_lens}. Superposed to the lensed field, we
display the cluster critical lines. Cluster members have been
identified by selecting galaxies with magnitude $m_{riz}<22$ and rms
radius $r_I>0.5"$. Comparing to the input catalogs of cluster members,
we find that foreground galaxies were efficiently identified by
applying these selection criteria. The images of the "test" galaxies
displayed in the left panel are indicated by letters followed by
numbers corresponding to their multiplicity. Note the radial image
labeled as G1 (one of the images of source G) is not detectable even
after subtraction of the foreground galaxies.

\subsection{Arc properties}
In some applications of strong cluster lensing, such as arc statistics,
it is important to accurately measure the properties of gravitational
arcs. Properties like length, width and curvature radius of the images
can be used to constrain cosmological parameters
\citep{BA98.2,BA03.1,WA03.1,OG03.1,LI05.1,ME04.1,HO05.1,HIL07.1} but also for determining the
inner-structure of galaxy clusters \citep{SA05.1}.  In this section we
illustrate a procedure to measure the shape of arcs that can be
applied to both real and simulated, thus favoring the comparison
between observations and simulations.

First, we identify the pixels in the image that belong to the lensed
images. To do that, we make local measurements of the sky background
and set a limit above which pixels can safely be assigned to the
arcs. For avoiding the inclusion of noise, especially where foreground
galaxies have been subtracted, we select only pixels above
$S/N>5$. The result of this selection is presented in the right panel
of Fig.~\ref{fig:cau_lens}. Several images cannot be classified as
arcs (for example the images C2, D2, etc), thus we exclude them from
the following analysis. We focus on the arcs A1, B1, C1, D1, E1 and
F1.

\begin{figure}[t]
  \includegraphics[width=\hsize]{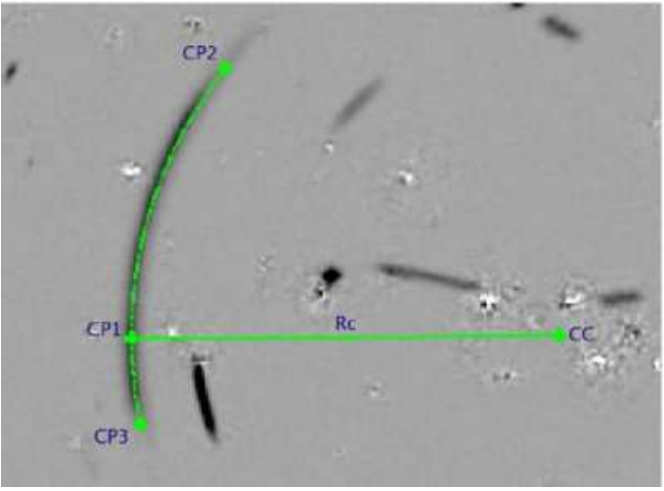}
  \caption{Fit of the arc A1: three characteristic points have been
    identified on the image (CP1-3), through which a circular segment
    (CP2-CP3) have been traced. The curvature radius (Rc) and the
    centre of curvature (CC) of the circular segment define the radius
    and the centre of the arc.}
\label{fig:a1}
\end{figure}

As suggested by \cite{MI93.1} \cite[see also][]{BA98.2,ME00.1,ME01.1},
we define three characteristic points on the arcs, namely 
\begin{enumerate}
\item the brightest point, that is likely to be an image of the source center;
\item the point at the largest distance from 1);
\item the point at the
largest distance from 2). 
\end{enumerate}
In the case of arc A1, these three points
are shown in Fig.\ref{fig:a1}. We trace a circle segment through these
characteristic points. Its length and curvature radius are identified
to the length and the curvature radius of the arc, respectively.

Measuring the arc width is less simple. Arcs are typically structured,
as they originate from mergers of multiple images of the same
source. Thus, even in the case of simple, elliptical sources, the
width is not constant along the arc. Moreover, arcs form in crowded
regions of clusters. As shown in the previous examples, the light from
cluster members can influence the detectability of the lensed images
or of part of them, thus affecting the measurements of the arc widths
and lengths. Cluster galaxies can also alter the shape of arcs by mean
of their lensing effects \citep{ME00.1}. Last but not less important,
photon noise causes the edges of the images to be irregular.

\begin{figure}[t]
  \includegraphics[width=\hsize]{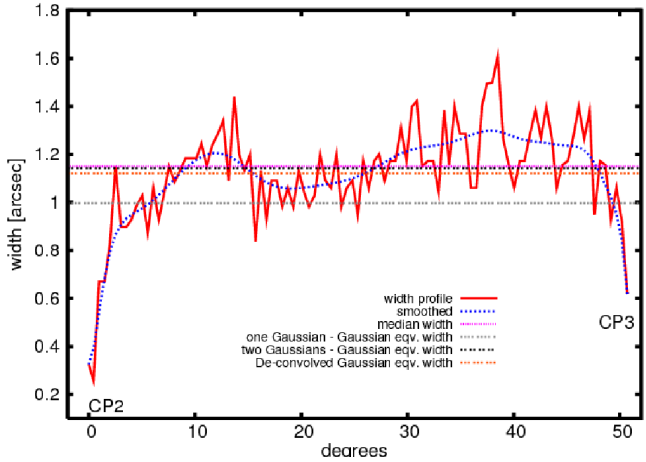}
\caption{Transversal width profile of arc A1 (red solid line). The blue dotted line shows an interpolation of the profile with a Bezier curve. The horizontal lines correspond to various determinations of the arc width.}
\label{fig:aza1}
\end{figure} 

The transversal width profile of arc A1 is shown in
Fig.~\ref{fig:aza1} (solid line). This has been recovered by making
radial scans of the arc along straight lines passing through the
center of curvature CC (see Fig.~\ref{fig:a1}) and intercepting the
circular segment CP2-CP3 at equi-spaced angular separations. The angle
is given in degrees with respect to the line CC-CP2. At each scan,
we measure the maximal distance between arc points intercepted by the
straight line. Some large scale fluctuations of the width can be
recognized, that are caused by mergers of multiple images. To make
these large scale modes more visible, we interpolate the profile with
a Bezier curve (dotted line) \citep{KN86.1}. The profile has several intervals with
positive curvature separated by valleys. The valleys are located at
the positions where the arc crosses the cluster critical line, as
shown in the central panel of Fig~\ref{fig:cau_lens}. They roughly
correspond to the regions where multiple images of the source have
merged together.  Additionally, many other small scale fluctuations
are produced by noise.  Thus, it appears difficult to define at which
position the arc width has to be measured. We propose here to use the
median width.

In the case of observations through the Earth atmosphere,
this effect is enhanced as both the width and the length of arcs are affected by the instrumental
PSF and by the seeing.  Since by definition arcs have large
length-to-width ratios, the blurring of the images produces a much
larger relative change of the widths than of the lengths. Thus, a
correction must be applied to the width measurements in particular.
This correction can be made by means of a de-convolution. Let assume
that the radial profile of the arc is well fitted by a Gaussian,
\begin{equation}
 	G(x)=Q \exp{-\frac{(x-\overline{x})^2}{2\sigma^2}} \;, 
\end{equation}
where $\bar x$ is the radial distance from the arc center of curvature. 
We define "Gaussian equivalent width" $W_G$ given by
\begin{equation}
	W_G \equiv 2 \sqrt{2 \ln{\frac{Q}{D}}} \sigma \;.
	\label{eq:gwidth}
\end{equation}
In this equation, $D$ represents the detection limit used to select
the arc points.

We assume that the total PSF (instrumental PSF+seeing) can be
described by another Gaussian function, whose FWHM is equivalent to
the PSF size:
\begin{eqnarray}
	G_{PSF} &=& \frac{1}{\sqrt{2\pi}\sigma_{PSF}}\exp{-\frac{x^2}{2\sigma_{PSF}^2}} \;, \\
	FWHM &=& 2\sqrt{2\ln{2}}\sigma_{PSF}\simeq 2.3548 \sigma_{PSF} \;.
\end{eqnarray}
We further assume that the intrinsic radial profile of the arc is also Gaussian:
\begin{equation}
 	G_0(x)=Q_0 \exp{-\frac{(x-\overline{x})^2}{2\sigma_0^2}} \;. 
\end{equation}
Then, the observed profile is a simple convolution of two Gaussians
and the parameters of the intrinsic profile can be derived from the
observed profile via the equations
\begin{eqnarray}
	\sigma_0&=&\sqrt{\sigma^2-\left(\frac{FWHM}{2.335}\right)^2} \;, \label{eq:sig0} \\
	Q_0&=&Q\frac{\sigma}{\sigma_0} \label{eq:Q0}\;.
\end{eqnarray}

Using Eq.~\ref{eq:gwidth}, the de-convolved Gaussian-equivalent width is 
\begin{equation}
	W_{G,0}=2 \sqrt{2 \ln{\frac{Q_0}{D}}} \sigma_0 \;.
	\label{eq:psfdecw}
\end{equation}

\begin{figure}[t]
  \includegraphics[width=\hsize]{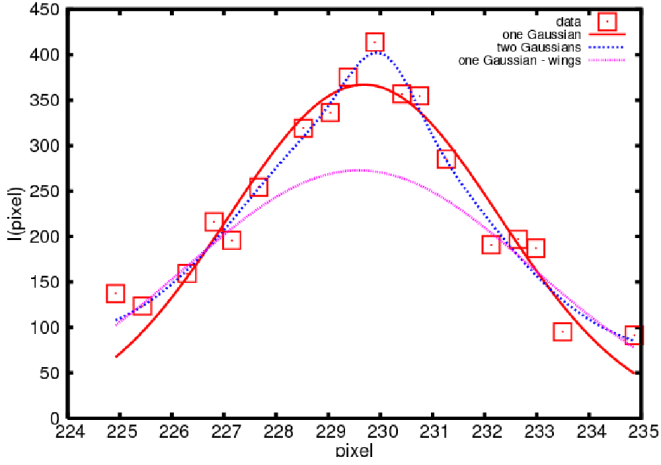}
  \caption{Median radial profile of arc A1 (square data
    points). Several attempts of fitting it are shown: single Gaussian
    (solid line), two Gaussians (long-dotted line) and single Gaussian
    fitting only the data points below half of the maximum of the
    profile (short-dotted line).}
\label{fig:radprof}
\end{figure}

Unfortunately, the radial profile of arcs is not well fitted by single
Gaussians. In Fig.~\ref{fig:radprof}, we show the median radial
profile of the arc A1. The data points are given by squares. The solid
line shows the best single Gaussian fit to the data. Clearly, a single
Gaussian provides a bad fit, both in the centre and in the wings of
the profile. In particular, in the wings the fit is lower than the
data. This implies that the Gaussian-equivalent width underestimates
the width of the arc. To be consistent with our definition of the
length, this is the maximal radial extension of the arc above a given
brightness threshold. In this case, the threshold corresponds to $\sim
100$ ADUs. At that level, the Gaussian-equivalent width is $\sim 2$
pixels (i.e. $\sim 0.23"$) less than the true width.

The fit improves if a combination of Gaussians are used:
\begin{equation}
 \tilde{G}(x)=\sum_{i=1}^M Q_i \exp{-\frac{(x-\overline{x}_i)^2}{2\sigma_i^2}} \;.
\end{equation} 
The coarsely-dotted line in Fig.~\ref{fig:radprof} shows the best fit
obtained by combining $M=2$ Gaussians. The reduced $\chi^2$ is smaller
by a factor of $\sim 2$ in this case. The resulting fit consists of a
combination of a broad Gaussian describing the wings of the profile,
and of a narrow one describing its central part and contributing very
little to the wings. This suggests that, in order to measure the width
of the arc, we could use a single Gaussian to fit the external parts
of the profile, without affecting significantly the measurements. For
example, the Gaussian that fits only the data points below the half of
the maximum of the profile is shown by the finely-dotted
line. Comparing to the previously described two-Gaussian fit, the
differences in the wings of the profile are minimal.

Even using multiple Gaussians, the de-convolution of the profile can
be done as described above. Indeed, each Gaussian component can be
de-convolved individually and summed up to provide a de-convolved
profile. The arc width can be then derived straightforwardly.

The same fitting procedure can also be applied to the tangential
profile of the arc, in order to correct for the PSF effects on the arc
length. However, as explained above, for arcs with large
length-to-width ratios, the correction is much less important than for
the arc width, since the tangential profile declines much more gently
than the transversal profile. For example, the best fit Gaussians
describing the decline of the tangential brightness profile on both
sides of arc A1 have $\sigma\gtrsim15$ pixels. The Gaussian best
fitting the wings of the radial profile shown in
Fig.~\ref{fig:radprof} has $\sigma=3.4$ pixels. Using
Eqs.~\ref{eq:gwidth} and \ref{eq:psfdecw}, the PSF correction amounts
to $\sim 2\%$ for the width and to $\lesssim 0.3\%$ for the length in
that example.

The width measurements made on arc A1 are displayed in
Fig.\ref{fig:aza1} using horizontal lines. Results are shown both for
the PSF convolved and de-convolved widths. The Gaussian-equivalent
width obtained by fitting the arc radial profile with a single
Gaussian is $0.99"$. This is $\sim 15\%$ smaller than the
Gaussian-equivalent width measured by using two Gaussians for fitting
the profile ($1.14"$). The latter is very similar to the true median
width ($1.15"$). Finally, the de-convolved width, obtained from the
two-Gaussians fit, is $1.12"$.

The analysis made on arc A1 has been repeated over the other arc-like
images in Fig.~\ref{fig:cau_lens}. We summarize the properties of all
these arcs in Table~\ref{tab:arcp}. In column 2 we list the arc
lengths (without PSF corrections), followed by several definitions of
arc widths, listed in columns 3-6. We report both the maximal and the
median widths and the Gaussian-equivalent widths before and after
PSF-correction. In column 7, we list the arc curvature radii.

The comparison between the median and the maximal widths shows that
large fluctuations of the arc thickness are possible along the
azimuthal profiles. Differences of order $30\%$ are typical but the
maximal width can be even $80\%$ larger than the median width, as
shown for the radial arc F1 (see table). This image is located in a noisy region, near to some cluster galaxies, thus the edges of the image are very
irregular. This confirms that defining the arc width - and other arc
properties that depend on this definition, like the length-to-width
ratio - is a critical issue.

The agreement between the Gaussian-equivalent and the median widths is
generally good. The PSF-corrections are typically of order few
percent, as expected given the narrow PSF of DUNE.

Some arcs are particularly straight. Two examples are the arcs D1 and F1 that have
very large curvature radii. In such cases, we arbitrarily assign a
curvature radius of $1000"$.

\begin{table}[t]

\begin{center}
\begin{tabular}{|c|c|c|c|c|c|c||c|}
\hline
\hline
{\rm arc} & $L$ & $W_{\rm max}$ & $W_{\rm med}$ & $W_G$ & $W_{G,0}$ & $R_c$ & $W_{0}$ \\
\hline
\hline
A1 & 23.2 & 1.61 & 1.15 & 1.14 & 1.12 & 26.2 & 1.13 \\ 
B1 & 14.3 & 0.98 & 0.81 & 0.81 & 0.77 & 31.5 & 0.7 \\ 
C1 & 4.55 & 0.88 & 0.73 & 0.76 & 0.7 & 70.6 & 0.69 \\ 
D1 & 3.91 & 0.83 & 0.62 & 0.61 & 0.59 & 1000 & 0.65 \\ 
E1 & 8.04 & 1.17 & 0.93 & 0.94 & 0.86 & 18.5 & 0.73 \\ 
F1 & 4.2 & 0.87 & 0.48 & 0.45 & 0.44 & 1000 & 0.47 \\ 
\hline
\hline
\end{tabular}
\end{center}
\caption{Summary of the main properties of the arcs shown in Fig.~\ref{fig:cau_lens}. Column 1: arc names. Column 2: arc lengths. Column 3-6: different definitions of arc widths (maximal, median, Gaussian-equivalent, PSF-corrected). Column 7: arc curvature radii. Column 8: arc widths in a simulation with null PSF. All quantities are given in arcsec.}
\label{tab:arcp}

\end{table}%

By activating and/or deactivating different features of the simulator,
the techniques used to analyze the images can be tested to quantify
how reliable are the measurements. In the last column of
Tab.~\ref{tab:arcp} we show the median widths measured in a simulation
where the PSF FWHM is set to zero. The remaining simulation parameters
(instrument, exposure time, etc.) are not changed. Although the PSF
correction is small, we notice that these newly determined widths are
in a slightly better agreement with the PSF-corrected widths in column
6 (r.m.s. deviation of $\sim 0.16"$) than with the previous median or
Gaussian-equivalent widths in column 4 and 5 (r.m.s. $\sim 0.23"$ and
$\sim 0.25"$, respectively). This suggests that the de-convolution
technique described above is correctly applied and allows to determine
the true widths with an uncertainty of order one pixel.

\subsection{Detectability of arcs}
An important question that can be answered using our simulator is how
do the properties of the arcs depend on source modeling.

First of all, we wish to determine for what intrinsic fluxes of the
sources are the arcs detectable in DUNE images, assuming an exposure time of $1000$ seconds.  For this purpose, we run several simulations,
using the same morphological models of the sources, i.e. without
changing their shapelet coefficients, their orientation and their
spectral type, but changing only the input fluxes such that the
apparent magnitudes in the $riz$ band vary between $m_{riz}=23$ and
$m_{riz}=28$. Then, we quantify if the arcs are detectable by counting
the number of pixels belonging to the arcs that emerge above a minimal
S/N ratio. As done before, we fix this limit to $S/N_{\rm min}=5$.

\begin{figure*}[ht]
  \includegraphics[width=0.5\hsize]{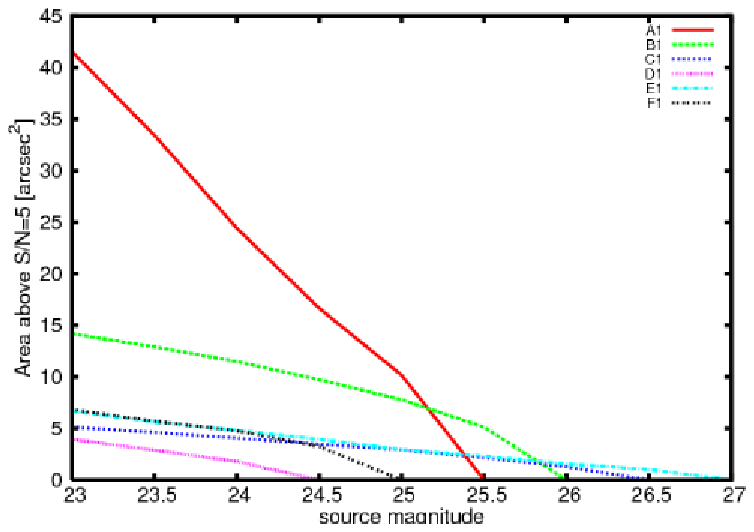}
  \includegraphics[width=0.5\hsize]{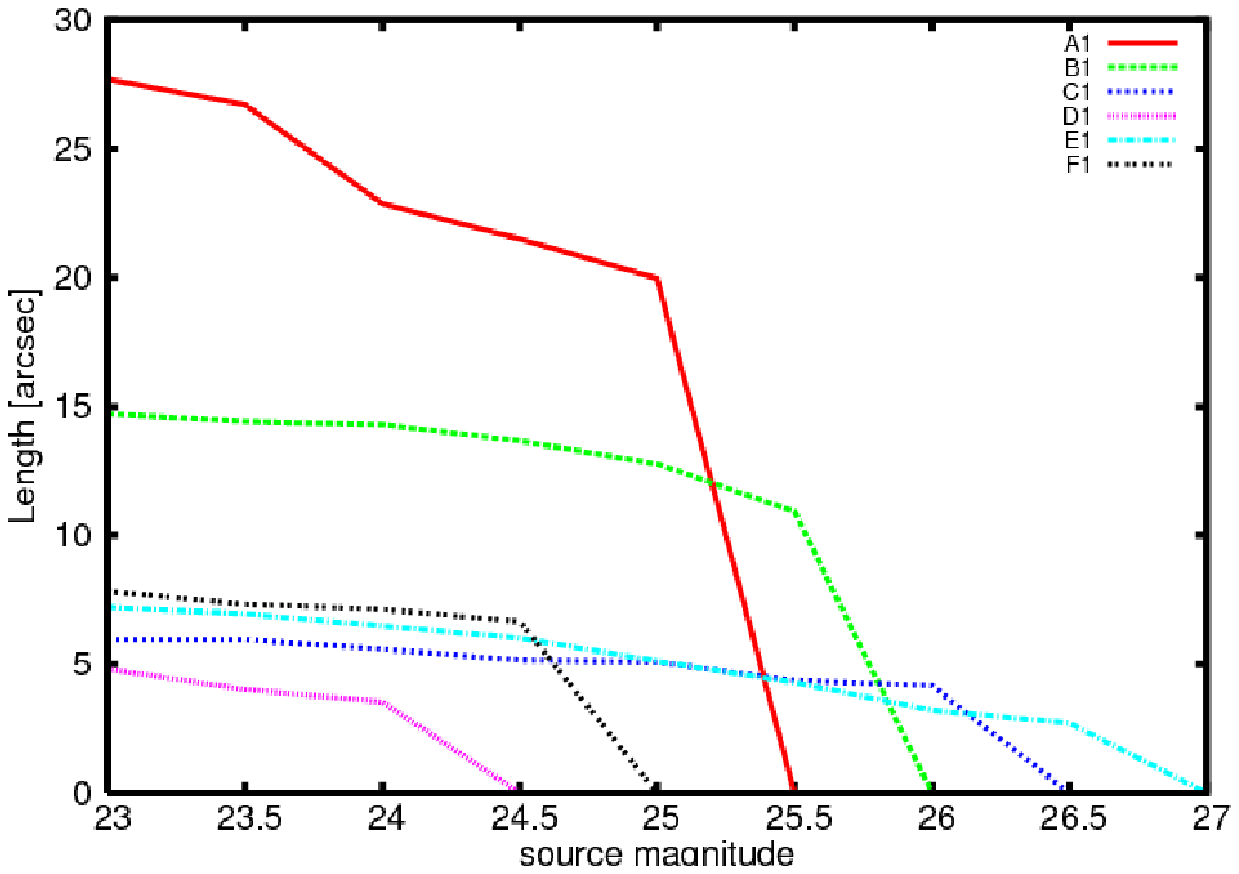}
  \includegraphics[width=0.5\hsize]{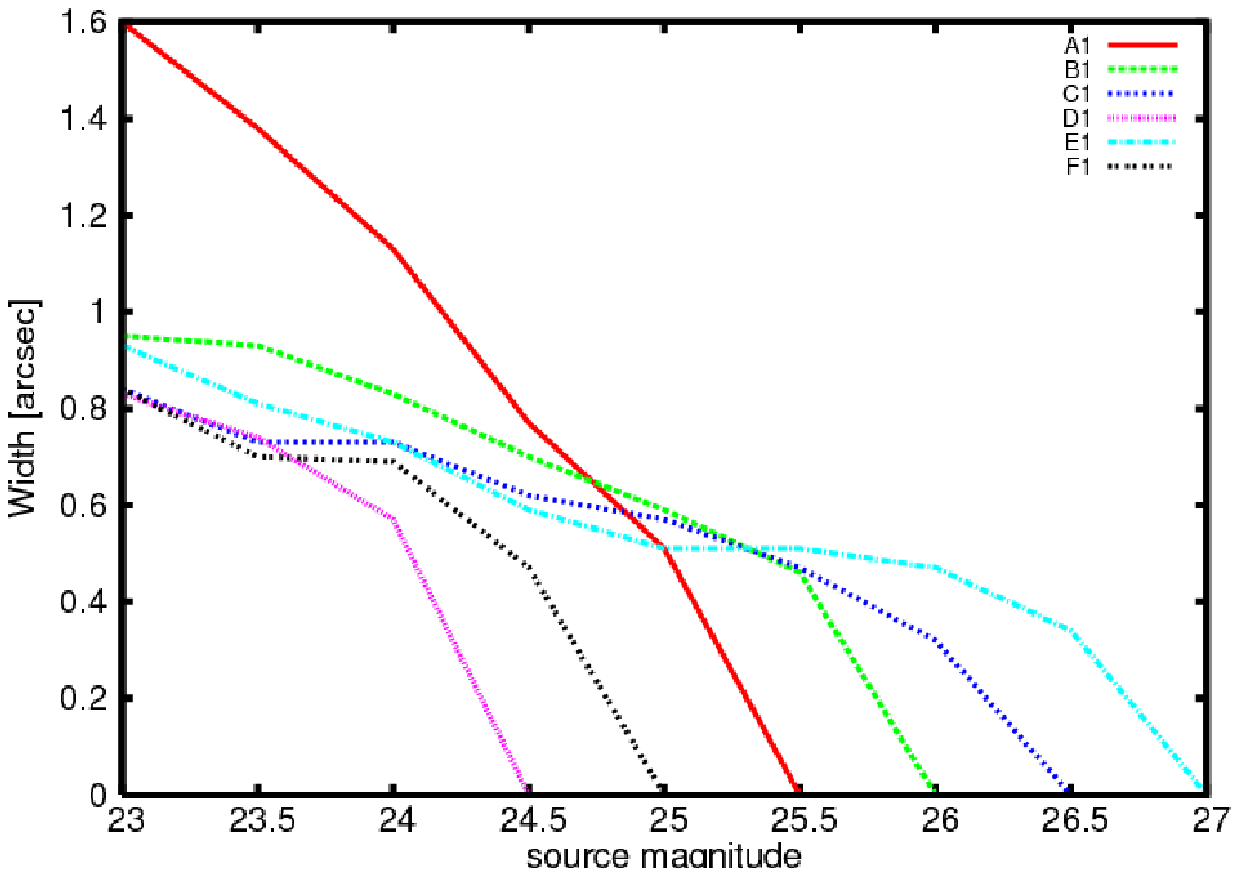}
  \includegraphics[width=0.5\hsize]{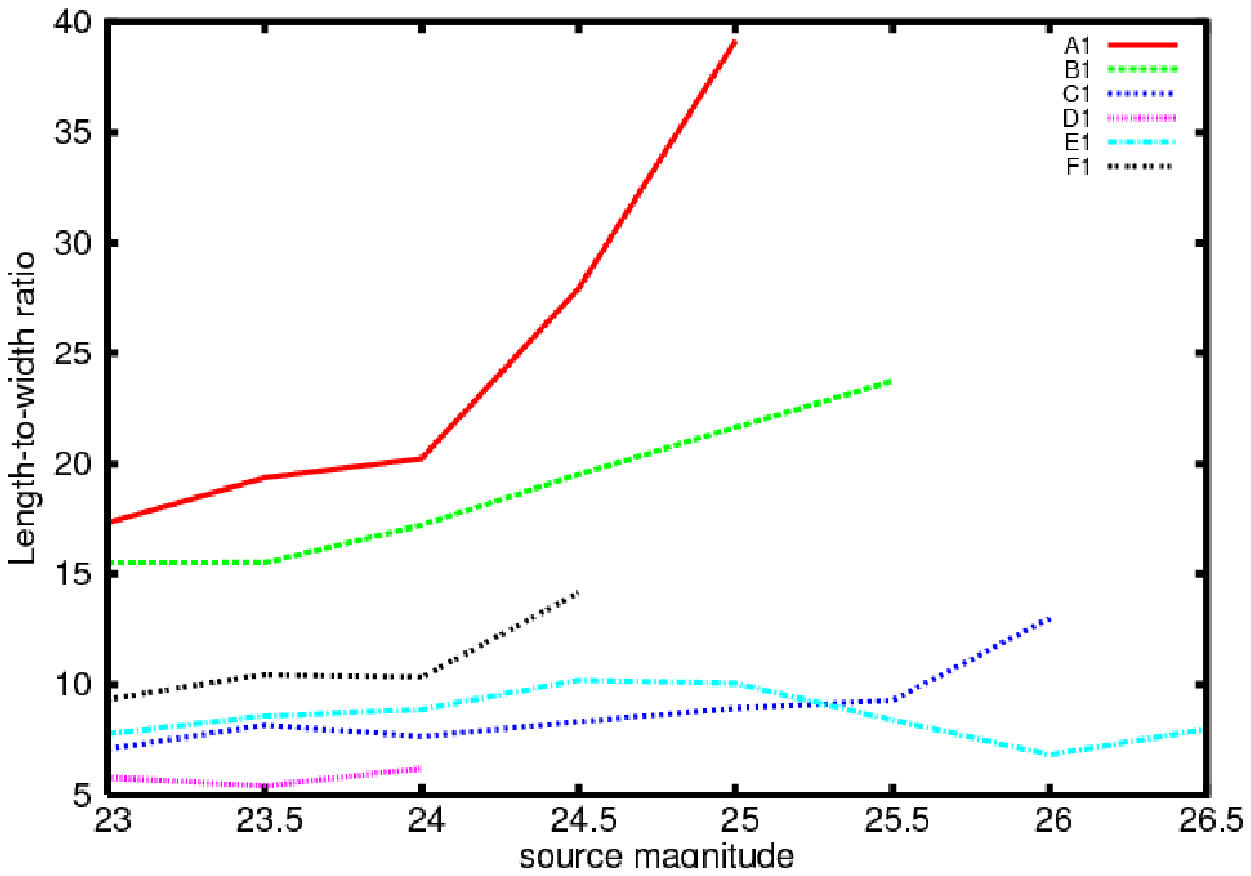}      
  \caption{Arc properties as a function of the source intrinsic
    magnitude in the $riz$ band. Shown are the area above the minimal
    signal-to-noise ratio $S/N=5$ (upper-left panel), the length
    (upper-right panel), the width (bottom-left) and the
    length-to-width ratio (bottom-right panel). The different
    line-styles correspond to the arcs A1-F1.}
\label{fig:propmag}
\end{figure*}

The results of this test are shown in Fig.~\ref{fig:propmag} for the
arcs $A1$ to $F1$ using different line-styles. In the top-left panel,
we show the arc areas above $S/N_{\rm min}$ as a function of the
intrinsic apparent magnitude of the sources (in absence of lensing
magnification). The limiting source magnitude for which arcs are
detectable significantly differs from arc to arc, ranging between
$24.5$ and $27$. The most prominent arcs, i.e. arcs A1 and B1, that
are the most magnified sources, can be detected only for source
intrinsic magnitudes below $25.5$ and $26$, respectively, while some
shorter arcs, like arcs C1 and E1, can be detected even if the sources
are fainter ($m_{riz}\lesssim 27$). We shall recall that lensing does
not change the surface brightness of the sources, but only the solid
angle under which they are observed. Thus, the differences between the
detectability limits must reflect the different size, shape and
brightness profile of the sources.
  
The observed properties of the arcs, in particular the length and the
width, obviously depend of the luminosity of the sources. In the
top-right and in the bottom-left panels of Fig.~\ref{fig:propmag} we
show how the lengths and the widths scale as a function of the
intrinsic magnitude. As expected, as the sources become fainter, the
arcs shrink in both the tangential and in the radial directions. Such
decrements, however, occur at different rates.  While the lengths
decrease slowly, the change of the arc widths is more rapid.  This is
clearly due to the presence of several bright knots along the
azimuthal profile of the arcs (see e.g. Fig.~\ref{fig:aza1}). These
originate from multiple images of the same source that merge from
opposite sides of the critical lines. Thus, as long as the arcs emerge
from the background, their length cannot become shorter than the
maximal separation between the knots. Only for very faint magnitudes
do the arcs break into smaller pieces and the lengths decrease
rapidly.  This behavior implies that the length-to-width ratio is not
a constant function of the source magnitude. As shown in the
bottom-right panel, especially for the longest arcs, the
length-to-width ratio is generally inversely proportional to the
source luminosity, once the source morphology has been fixed. For
example, between $m_{riz}=23$ and $m_{riz}=25$, the length-to-width
ratio of the arc A1 becomes larger by more than a factor of two. For
arc B1, it grows by $\sim 65\%$ between $m_{riz}=23$ and
$m_{riz}=25.5$. Also the length-to-width ratios of the arc C1 and of
the radial arc F1 grow significantly.

\begin{figure}[t]
  \includegraphics[width=\hsize]{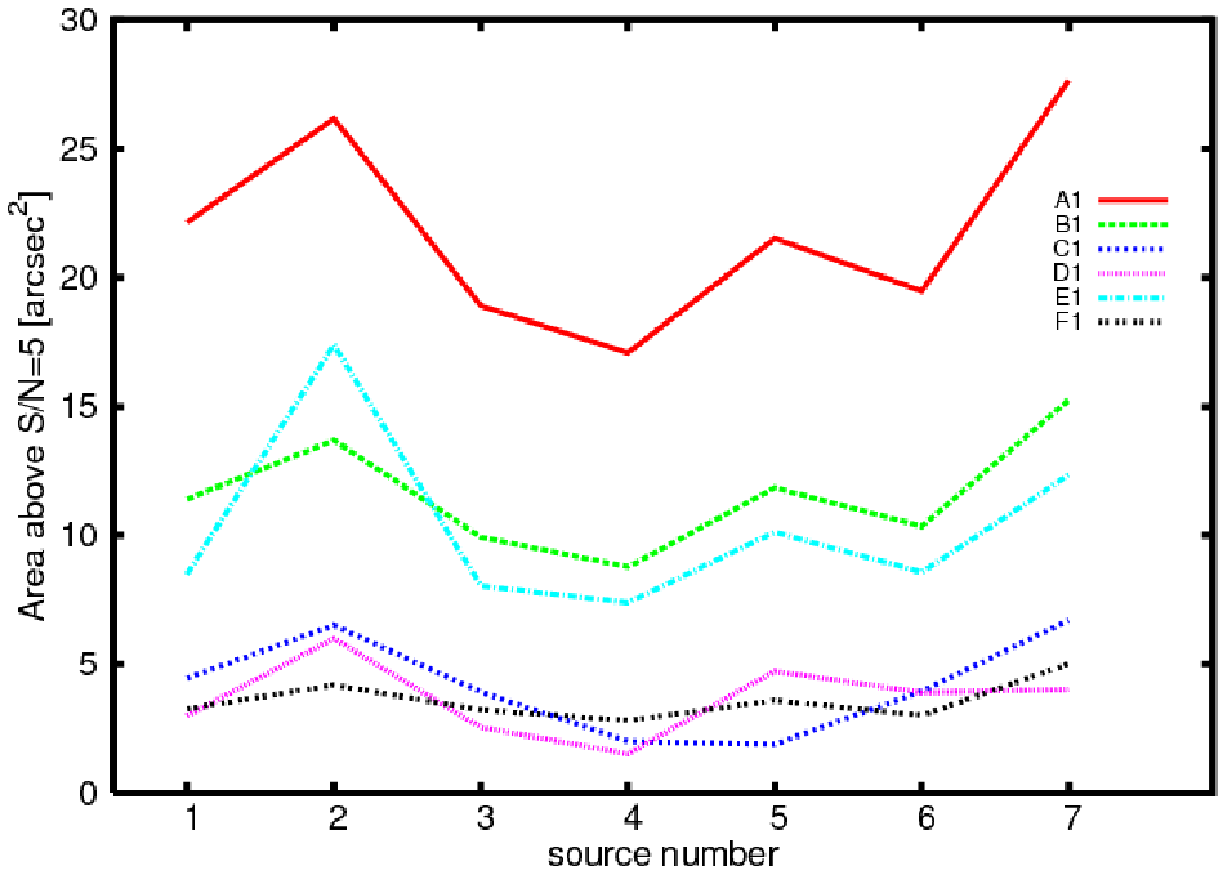}
  \includegraphics[width=\hsize]{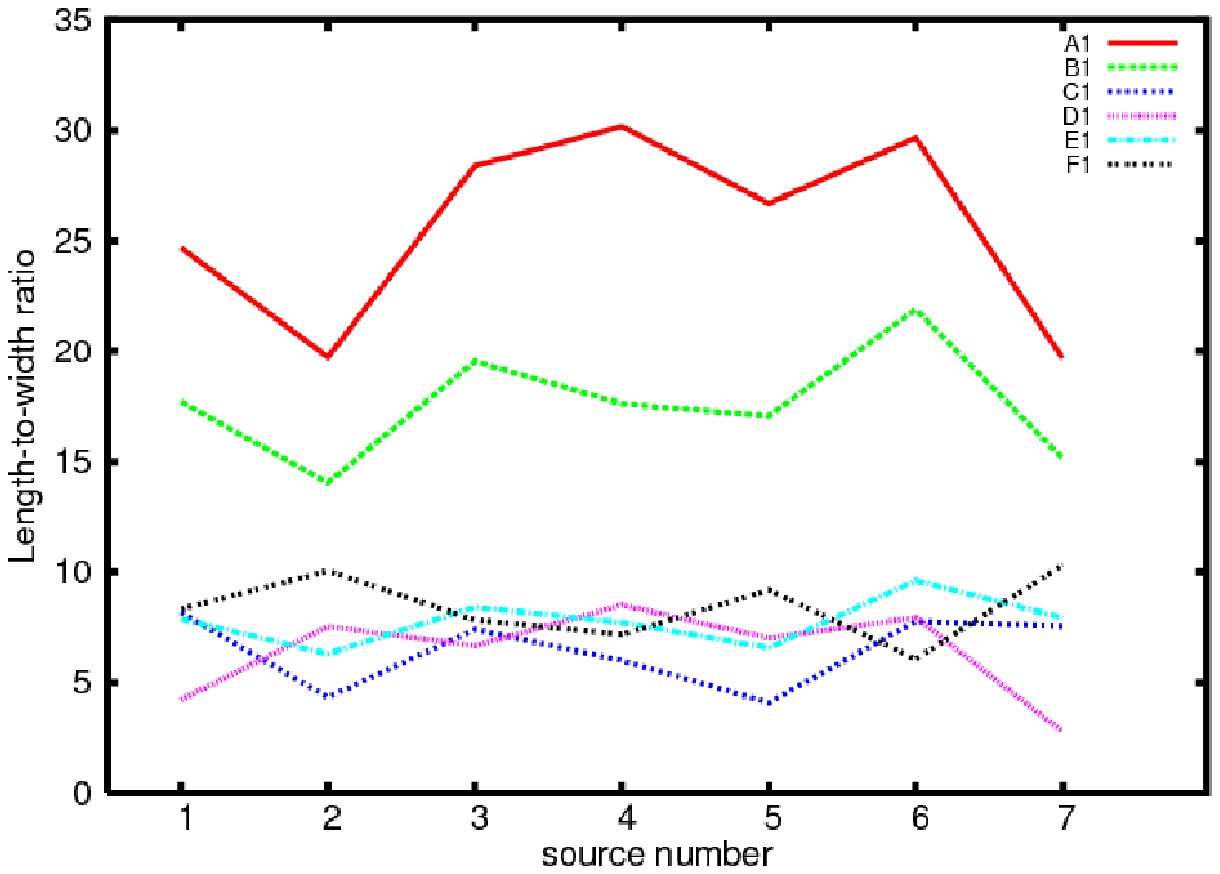}
  \includegraphics[width=0.137\hsize]{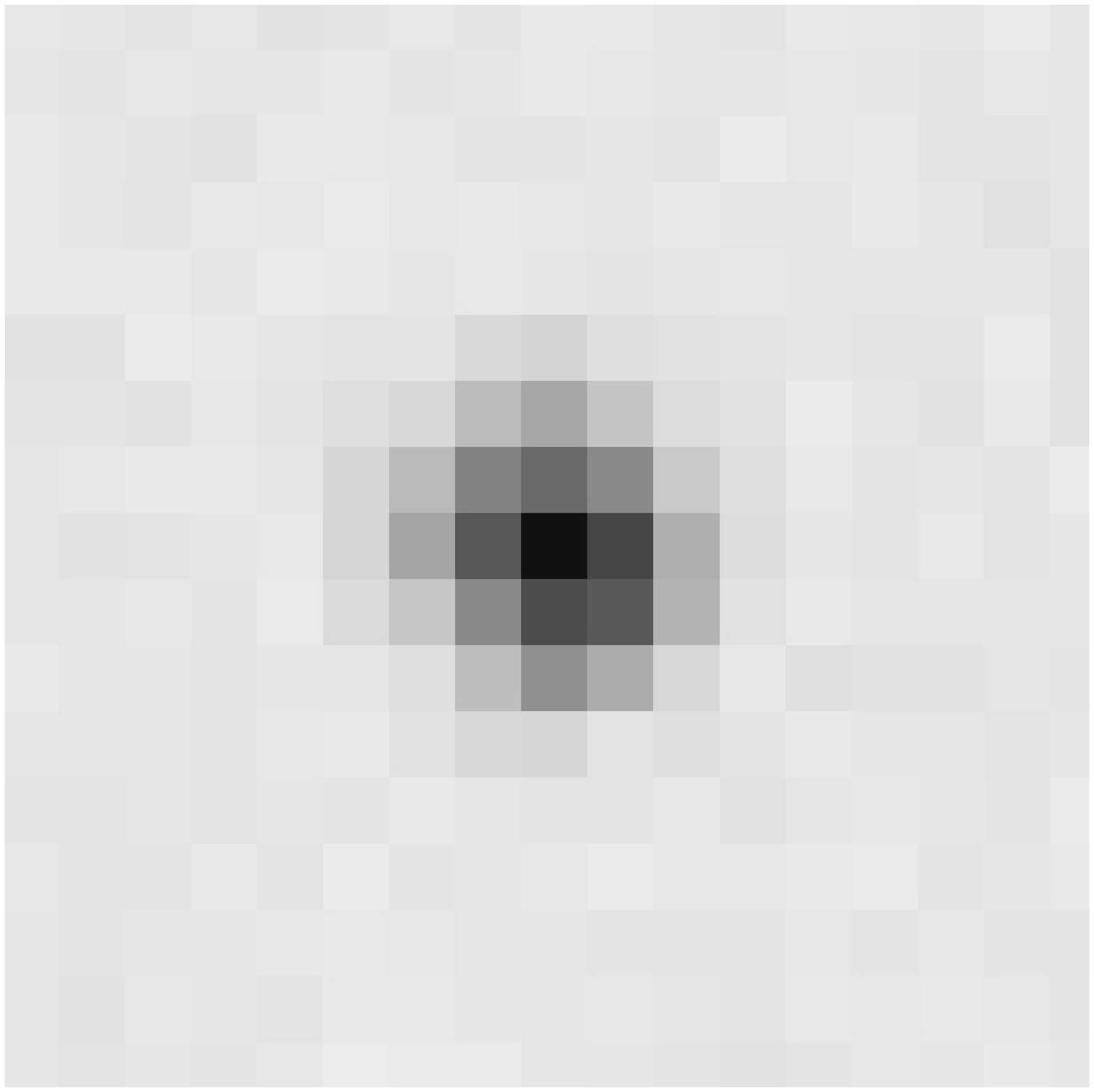}
  \includegraphics[width=0.137\hsize]{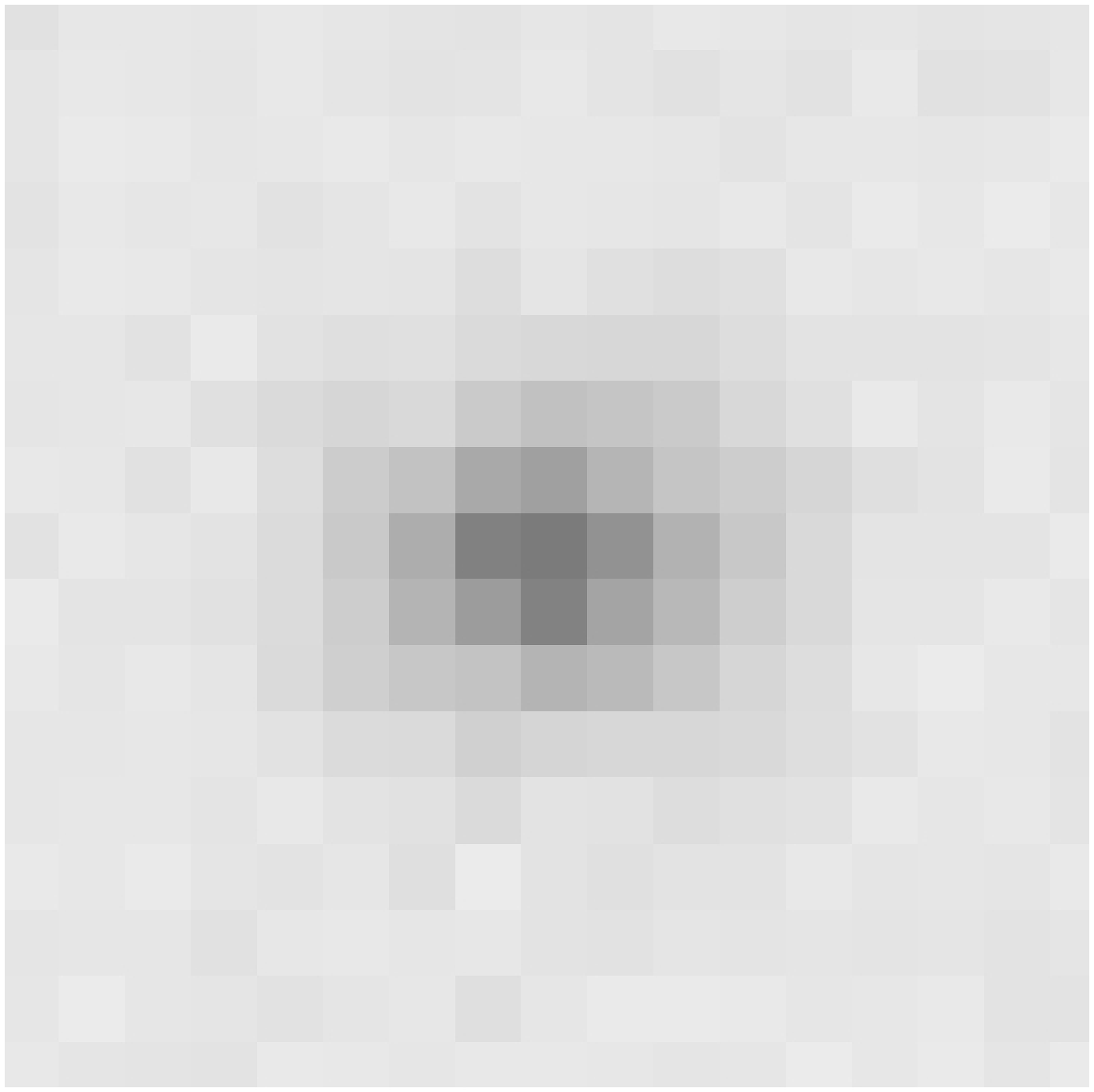}
  \includegraphics[width=0.137\hsize]{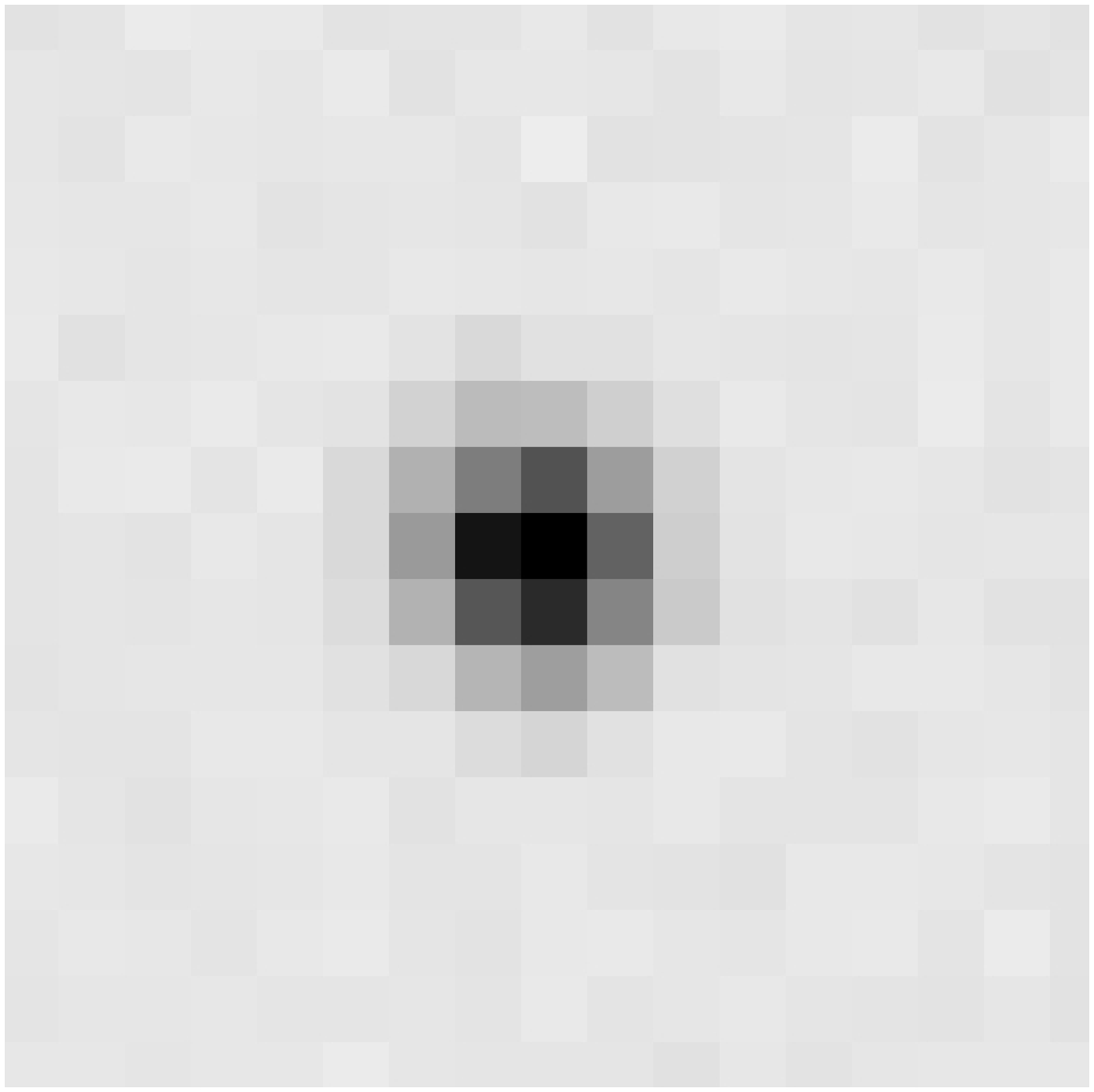}
  \includegraphics[width=0.137\hsize]{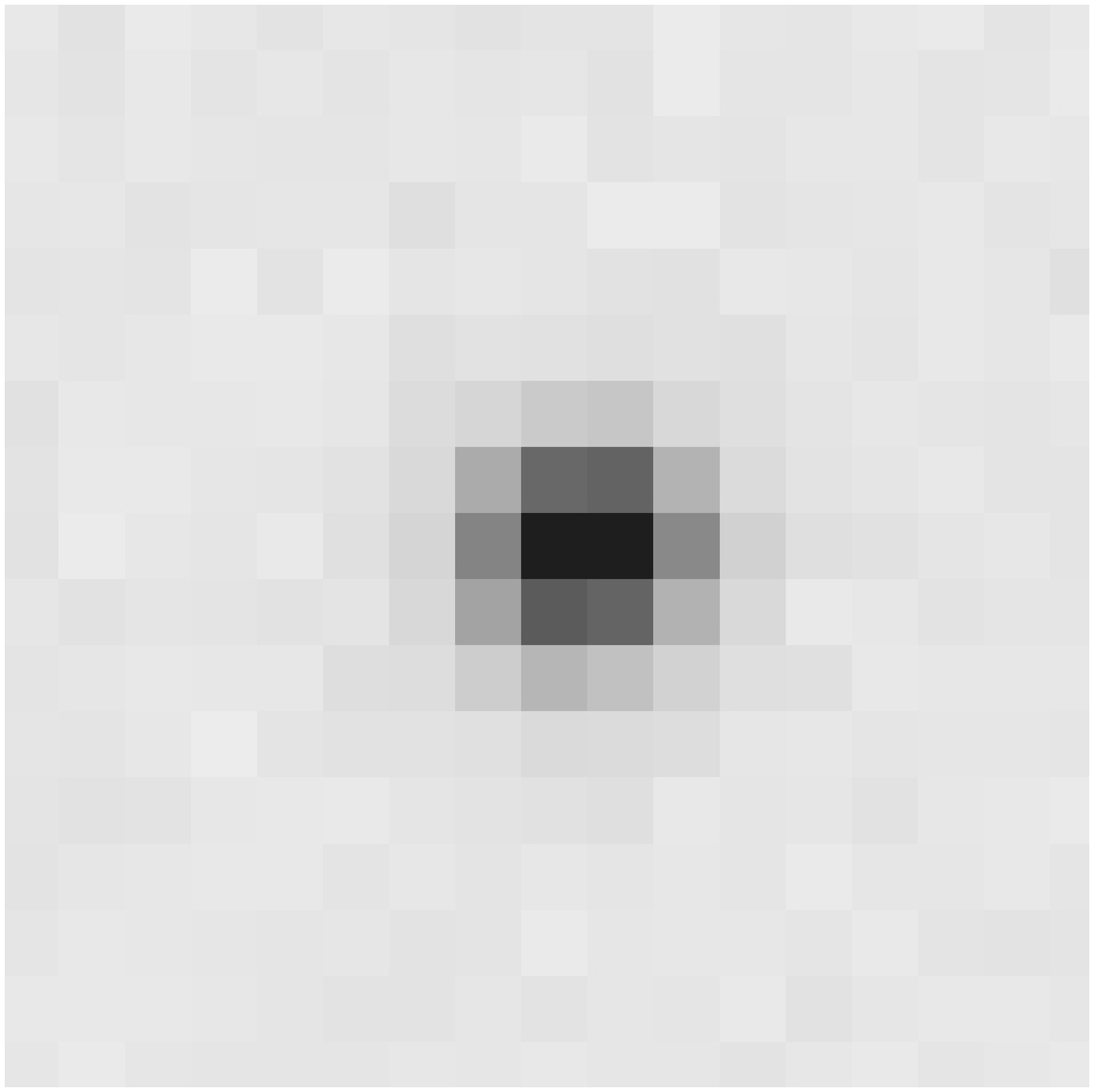}
  \includegraphics[width=0.137\hsize]{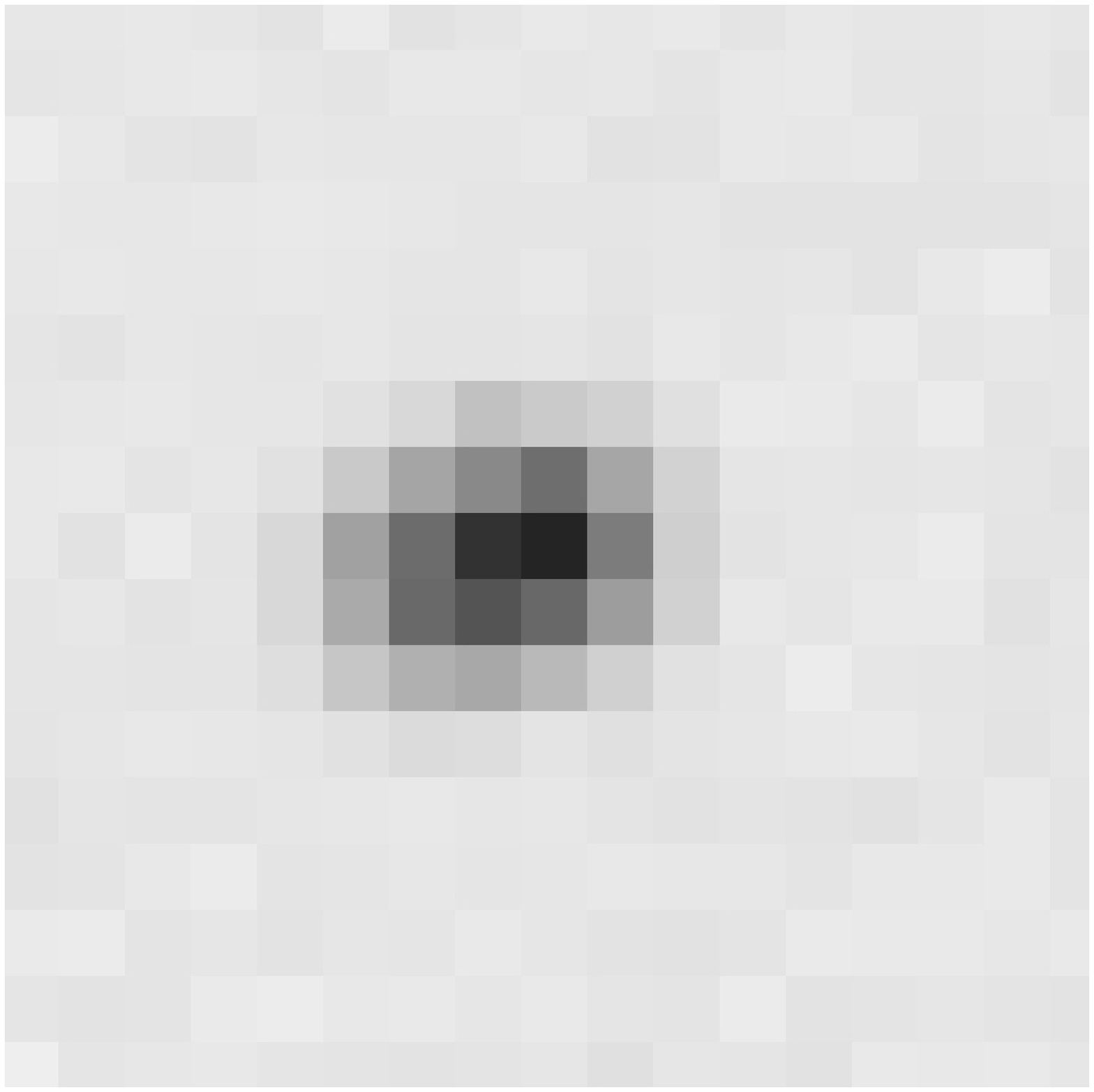} 
  \includegraphics[width=0.137\hsize]{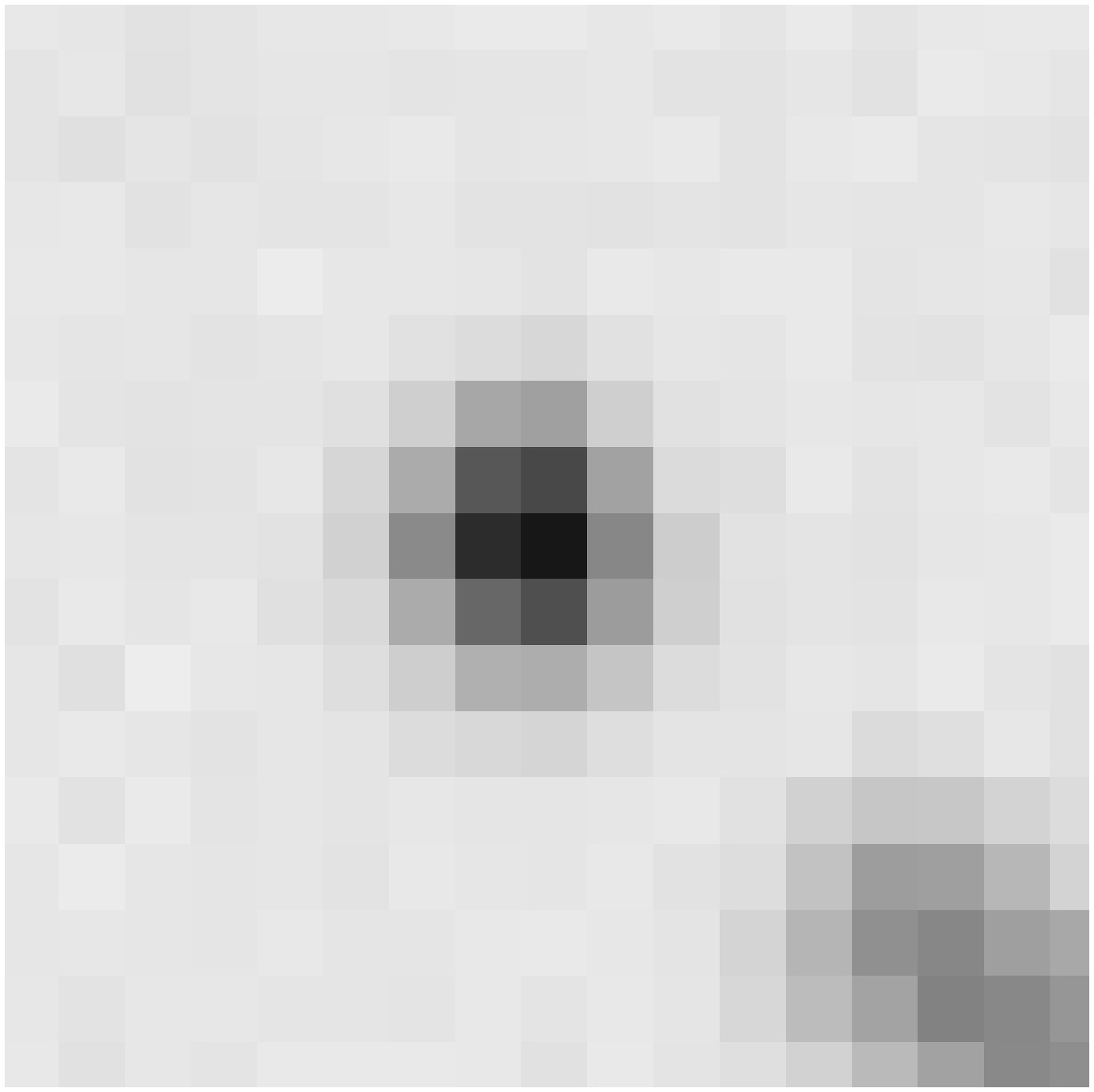}
  \includegraphics[width=0.137\hsize]{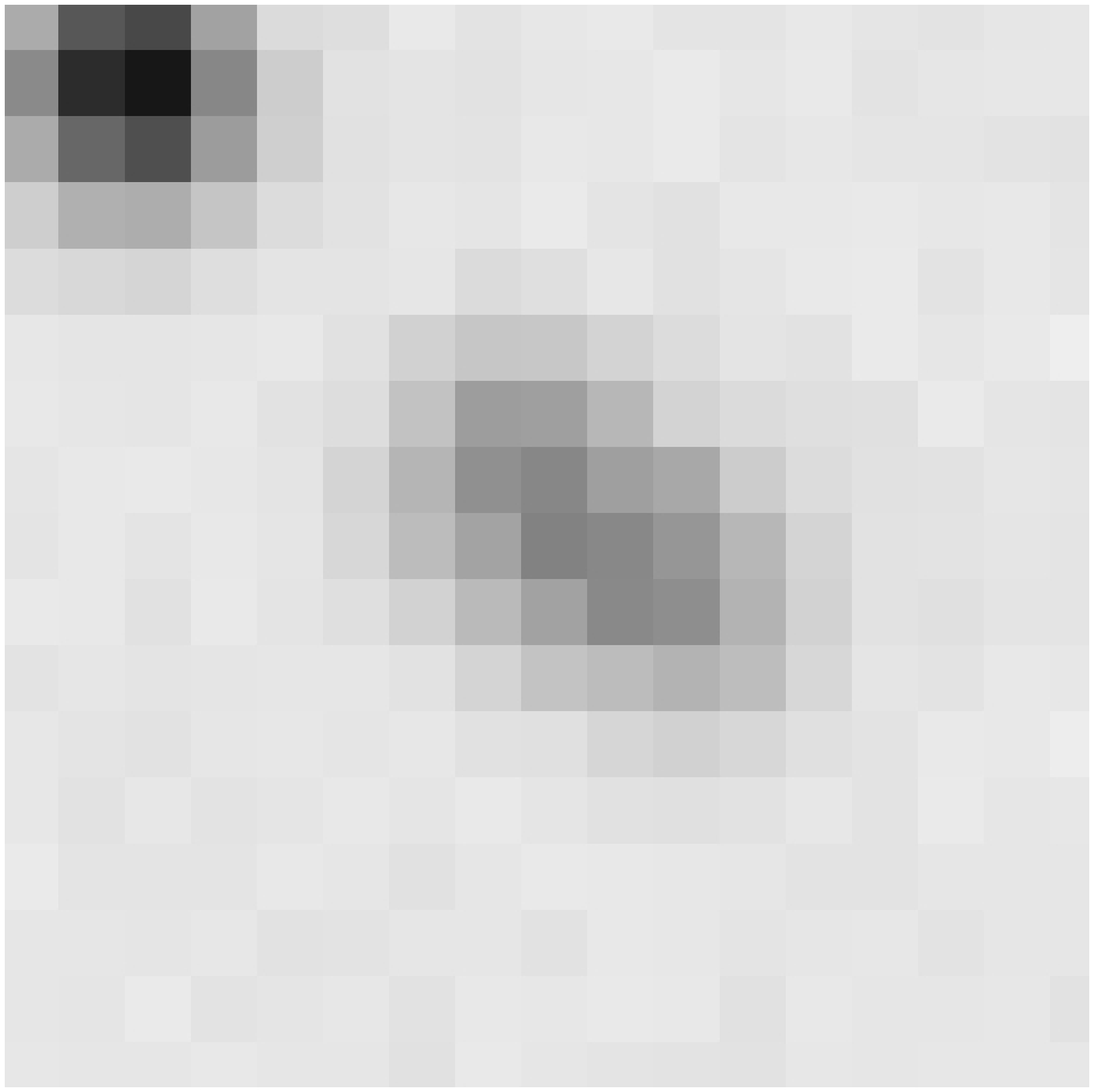}           
  \caption{Top panel: Areas of the arcs A1-F1 above $S/N=5$, assuming
    different sets of shapelet coefficients for modeling the
    sources. The source models (shown in the figures below the graphs)
    are numbered from 1 to 7. All sources have intrinsic magnitude
    $m_{riz}=24$. Bottom panel: length-to-width ratio as a function of
    the source model}
\label{fig:propsource}
\end{figure}

To better explain the dependence of the arc appearance on the
intrinsic properties of the sources, we run new simulations, keeping
the sources at the same positions but exchanging among them the sets
of shapelet coefficients that determine their morphology and surface
brightness distributions. In other words, we re-simulate each arc
using seven different source shapes. The intrinsic magnitude of the
sources was fixed to $m_{riz}=24$. This value has been chosen such
that all arcs A1-F1 are detectable. In the upper graph of
Fig.~\ref{fig:propsource} we show the area of each arc above $S/N=5$
as a function of the source model. Each model is identified by a
number running from 1 to 7. The corresponding un-lensed images of the
sources are shown at the center of the frames below the graphs. Note that these are cut-outs from the left panel of Fig.~\ref{fig:cau_lens}. As anticipated earlier, by varying
the morphology of the sources, different amounts of pixels belonging
to the arcs surpass the detectability thresholds. In particular, the
spatial extent of the arc is determined by the size of the sources,
i.e. by their surface brightness. We notice that the curves
corresponding to different arcs have many features in common. Indeed,
they have several local minima and maxima that arise for the same
source models. For example, all arcs have the largest sizes if the
source model number 2 or number 7 are used, while their size is the
smallest for the source model number 4. Looking at the corresponding
source representations, we see that the former are low
surface-brightness and spatially extended galaxies, while the latter
corresponds to the most compact among the sources.
        
We now consider the correlation between the source model and the
length-to-width ratio of their images. This is shown in the bottom
graph of Fig.~\ref{fig:propsource}. Compared to the previous plot, we
notice the opposite correlation for most of the images: for a fixed
total luminosity of the sources, the most extended galaxies (the sources 2 and 7 in Fig.~\ref{fig:propsource}) correspond
to the lowest arc length-to-width ratios.

We can interpret these results as follows. Large, low surface
brightness images average magnification over a larger area. Thus, they
are less distorted. Indeed, the relative changes of the lengths and of
the widths of the images are caused by the convolution of the
tangential and of the radial magnifications with source surface
brightness distribution. The distortion effects are diluted if the
source is broader. For this reason, arcs originating from compact,
faint or low surface brightness sources that are barely detectable are
preferentially characterized by large length-to-width ratios. These results show that comparing theory to observations, we need to know the source population the arcs are drawn. As surface brightness is unchanged, we suggest basing the detectability criterion of arcs upon it.

\section{Conclusions}
\label{sect:conclu}

We presented a new code for mimicking optical observations of distant
galaxies. It includes all relevant sources of noise affecting real
observations, such as sky background, photon noise and effects of the
atmosphere and of the instrumental PSF. The code also incorporates
ray-tracing routines that allow us to include lensing effects by matter
distributed between the observer and the sources.

In this work we have detailed the
thorough method that we have developed in order to model the source
galaxies realistically. The
source morphologies are modeled via the shapelet decomposition of real
galaxy images observed with HST. Four spectral templates are used to
simulate the spectral energy distributions of the galaxies. Their
redshift and luminosity distributions as a function of their spectral
classification are derived from observed galaxy luminosity functions
in the VVDS.

Being particularly suitable for simulating lensing by galaxy clusters,
the code also simulates the emission from the cluster galaxy
population, whose properties are obtained from semi-analytic models of
galaxy formation and evolution.

Many features of the simulator have been discussed in detail. In
particular, we have shown that the code can be easily used to mimic
observations with a variety of existing instruments, both from the
space (HST) and from the ground (LBT).
 
In order to illustrate the code and to introduce some potential
applications, we have simulated observations of a
galaxy cluster obtained from N-body simulations. These virtual
observations were made using a telescope with the characteristics of
the proposed space telescope DUNE.

The simulated images were analyzed using standard observational
techniques. After subtracting the light of the foreground galaxies,
multiple images of several distant sources that were strongly lensed
by the foreground cluster can be identified. Each detected arc-like
image has been classified, measuring its length, width and curvature
radius.

We discussed how the properties of arcs can be measured consistently
in simulations and observations, in order to facilitate the comparison
with theoretical predictions.  In particular, we focused on the
determination of the arc widths and on how they can be corrected for
the PSF broadening. For this purpose, we fit the radial profile of the
arcs with multiple Gaussians. By comparing simulations with and
without PSF convolution, we verified that the de-convolution technique
allows one to determine the true width of the arcs with a typical error of
one pixel.

By varying the characteristics of the sources, we studied the
detectability limits of arcs in DUNE observations. With an exposure
time of $1500$ seconds, we expect that DUNE will be able to observe
arcs arising from sources whose magnitudes are fainter than $\sim 27$ in
the $riz$ band. The shape of the arcs are found to be very sensitive
to the properties of the sources. In particular, we found that arcs
tend to acquire larger length-to-width ratios as their sources become
fainter or more compact.

These results are particularly intriguing for arc statistics. Indeed,
they show that observational effects may have a large impact on the
abundance of arcs with large length-to width ratios, and they deserve
careful investigation. We will dedicate a forthcoming paper to this
subject (Meneghetti et al., in prep.). The source morphologies, and in particular sub-structures of star-forming
regions, are clearly more relevant for arc statistics than is usually
assumed.  This is as a result of the small angular size of the bright star
forming regions which are more easily distorted. Moreover, the number of very elongated images is expected to
grow rapidly as a function of the depth of the observations. Thus,
space missions like DUNE, capable of making deep and wide surveys
thanks to their large sensitivity and field of view, are likely to
discover a significant number of new gravitational arcs. The usage of
efficient software for automatic arc identification \citep[see
e.g.][]{LE05.1,SE07.1,AL06.1,HO05.1,CA07.1} will facilitate the
detection of strong lensing events in these large data sets.

\acknowledgements{The $N$-body simulations were performed at the
  ''Centro Interuniversitario del Nord-Est per il Calcolo
  Elettronico'' (CINECA, Bologna), with CPU time assigned under an
  INAF-CINECA grant. This work has been supported by the Vigoni
  programme of the German Academic Exchange Service (DAAD) and
  Conference of Italian University Rectors (CRUI). We acknoweldge
  financial contribution from contracts ASI-INAF I/023/05/0, ASI-INAF
  I/088/06/0 and INFN PD51. CH acknowledges the support of a European Commission Programme 6th
framework Marie Curie Outgoing International Fellowship, under contract
M01F-CT-2006-21891.
 MM thanks M. Bolzonella, G. Rodighiero,
  E. Zucca, S. Bardelli, E. Vanzella and G. Zamorani for helpful
  discussions. We thank A. Refregier, J. Rhodes and the whole DUNE
  collaboration for allowing us to test our code during the preparation of the
  DUNE proposal. We thank M. Barden and S. Koposov, the GOODS and COMBO-17 team for providing the GOODS galaxy images used for the presented work. We are grateful to C. Peng for providing the software
  GALFIT and to C. Wolf and K. Meisenheimer for giving us the SEDs of the GOODS galaxies. We warmly thank V. Springel for providing us with
  post-processing software. MM thanks Paolo for his cooperation during the preparation of the manuscript.}

\bibliography{./TeXMacro/master}
\bibliographystyle{aa}

\end{document}